\documentclass{pnastwo}
\usepackage{sidecap, epsfig, amsmath,amssymb,amsfonts,graphicx}
\usepackage{pnastwof, caption}

\newcommand{\superscript}[1]{\ensuremath{^{\textrm{#1}}}}

=\frutigeroblique at 8 pt
=\frutiger at 7 pt
=\frutigerbold at 7 pt
=\frutigerboldoblique at 7.5 pt

\newcommand{\boldS}{\mbox{\boldmath$S$}}
\newcommand{\boldnu}{\mbox{\boldmath$\nu$}}
\newcommand{\boldw}{\mbox{\boldmath$w$}}

\begin{document}
\pagestyle{plain}
\title{Dispensability of {{\helveticaboldoblique at 22 pt \it Escherichia coli\,'}}s latent pathways}

\author{
Sean P. Cornelius
\affil{1}{Department of Physics and Astronomy, Northwestern University, Evanston, IL 60208, USA},
Joo Sang Lee
\affil{1}{},
\and
Adilson E. Motter
\affil{1}{}
\affil{2}{Northwestern Institute on Complex Systems, Northwestern University, Evanston, IL 60208, USA}
\affil{3}{Department of Molecular Biology, Princeton University, Princeton, NJ 08544, USA}
}

\contributor{Published in PNAS 108, 3124--3129 (2011)}

\maketitle

\begin{article}

\begin{abstract}
Gene-knockout experiments on single-cell organisms have established that
expression of a substantial fraction of genes is not needed for optimal growth. This problem acquired
a new dimension with the recent discovery that environmental and genetic
perturbations of the bacterium \textit{Escherichia coli} are followed by the
temporary activation of a large number of latent metabolic pathways, which
suggests the hypothesis that temporarily activated reactions impact growth and
hence facilitate adaptation in the presence of perturbations. Here we test this
hypothesis computationally and find, surprisingly, that the availability of
latent pathways consistently offers no growth advantage, and tends in fact to
inhibit growth after genetic perturbations. This is shown to be true even for
latent pathways with a known function in alternate conditions, thus extending
the significance of this adverse effect beyond apparently nonessential genes.
These findings raise the possibility that latent pathway activation is in fact
derivative of another, potentially suboptimal, adaptive response.
\end{abstract}


\dropcap{L}iving cells are surprisingly robust against mutations and, in
particular, against gene knockouts
\cite{baba_construction_2006,gerdes_experimental_2003,giaever_functional_2002,
deluna_exposingfitness_2008,delia_are_2009}.  The origin of mutational
robustness---whether it is a directly evolved trait or a byproduct of
evolutionary history---remains debatable~\cite{kupiec_genetic_2007}. In either
case, metabolic network analysis shows that the nonessentiality of enzymes and
associated genes is largely due to the inactivity of the corresponding metabolic
reactions under laboratory
conditions~\cite{papp_metabolic_2004, nishikawa_spontaneous_2008,
blank_large-scale_2005}. This leaves environmental robustness as the natural
candidate to explain gene nonessentiality. Yet, apart from chemical stress-based
assays~\cite{hillenmeyer_chemical_2008}, studies designed to test whether
nonessential genes become essential under different conditions have
failed to identify a phenotype for more than a small fraction of
additional genes~\cite{harrison_plasticity_2007}. A recent groundbreaking study
has shown, however, that a large fraction of reactions not active under standard
laboratory conditions become transiently active after a genetic or environmental
perturbation~\cite{fong_latent_2006,fong_parallel_2005}. Why? The prevailing
interpretation has been that the transient activation of such latent pathways
facilitates adaptation to new conditions, thereby attributing  function to genes
that have been classified as dispensable for the lack of phenotype  in
steady-state experiments. This is naturally formulated as the hypothesis that
latent pathways have a positive impact on postperturbation growth (cellular
reproduction), which is a measure of competitive advantage with a strong
empirical basis \cite{baba_construction_2006, gerdes_experimental_2003,
giaever_functional_2002, hillenmeyer_chemical_2008, harrison_plasticity_2007,
fong_parallel_2005}. Even for genes with known functions under different
conditions, this hypothesis is appealing as it suggests the possibility of an
alternate phenotype that would not be detected in traditional high-throughput
screens of knockout mutants \cite{ito_functional_2005}.

Here we test this hypothesis using the most complete in silico
reconstruction of the metabolic network of \textit{Escherichia coli} K-12 MG1655
\cite{feist_genome-scale_2007,feist_growing_2008} and perturbations caused by
single-gene knockouts. The response of the metabolic network to knockout
perturbations is modeled using both model-independent analysis and the two most
accepted phenomenological approaches, minimization of metabolic adjustment
(MOMA) \cite{segre_analysis_2002} and regulatory on/off minimization (ROOM)
\cite{shlomi_regulatory/off_2005} ({\it Materials and Methods}). Starting from a
growth-maximizing state determined by flux balance analysis (FBA)
\cite{edwards_escherichia_2000}, we compare the early postperturbation growth
rate ({\it Materials and Methods}) of the original organism with that of a modified
organism in which the latent reactions have been disabled. We consider glucose
minimal medium and gene knockouts that necessarily change the original metabolic
flux distribution but that nonetheless are compatible with nonzero growth
according to FBA. There are 52 enzyme-coding genes associated with 97 metabolic
reactions in the reconstructed network that satisfy this condition. We
systematically predict the impact of latent pathway activation on growth rate
following perturbations caused by the knockouts of each of these genes.

\section*{Results}
\subsection*{Phenomenological analysis.} \ Figure 1 illustrates the
essence of our approach. In an optimal growth state, as observed for \textit{E.
coli} after adaptive evolution in fixed environmental conditions
\cite{ibarra_escherichia_2002}, many metabolic reactions are inactive
\cite{nishikawa_spontaneous_2008,fong_latent_2006}. Shortly after a
perturbation, however, both the original and modified strain operate in a
suboptimal growth state \cite{segre_analysis_2002,shlomi_regulatory/off_2005},
which we model using MOMA or ROOM (Fig. 1{\it A}).  In silico
\cite{nishikawa_spontaneous_2008} and laboratory \cite{fong_latent_2006}
experiments show that this change is accompanied by a burst of reaction activity
(Fig. 1{\it B}), reflecting regulatory changes that locally reroute
fluxes in the short-term metabolic response to the perturbation
\cite{emmerling_rerouting_2002, huang_rerouting_2003}. If the perturbation is
nonlethal, the perturbed organisms will undergo adaptive evolution---adopting
beneficial mutations over longer timescales \cite{herring_2006,
lee_adaptive_2010} to achieve a new optimal growth state, which can be predicted
by FBA \cite{fong_metabolic_2004, shlomi_regulatory/off_2005}.

For the perturbations considered in this study, the average and standard
deviation of the number of transiently active reactions is $291 \pm 83$ and $120
\pm 59$ for MOMA and ROOM, respectively. This difference is expected since ROOM,
by design, favors a small number of significant flux changes, which reflects the
fact that ROOM may model a later stage in the adaptive response pictured in
Fig. 1{\it B} than MOMA \cite{shlomi_regulatory/off_2005}. These numbers should
be compared with the number of active reactions in the corresponding
growth-maximizing states, which is $385$ in the wild type and remains $385$ on
average in the knockout mutants, with an average of $\approx$ $98$\% overlap
between the two sets for the simplex solutions we consider; these numbers are
representative for other choices of optima \cite{mahadevan_effects_2003,
reed_genome-scale_2004} within our models (Supporting Information, {\it
Sensitivity to Alternate Optima} section). We emphasize that since the modified organism
lacks only transiently active (latent) metabolic reactions, the
optimal steady states are identical to those of the unmodified strain both
before and after the perturbation. Our question is then whether the early
postperturbation growth rate (before adaptive evolution) will be lower (red), remain equal 
(green), or become higher (blue) when these latent
pathways are not present (Fig. 1{\it A}).

Our principal result is that the strains lacking latent pathways systematically
show equal or better adaptation as determined by growth within our in
silico model, regardless of the approach used to model the organisms' response
to perturbations. We assume that the organisms are in an optimal growth state
both before and long after the perturbation, which accounts for cases that have
received much attention in the literature \cite{nishikawa_spontaneous_2008,
fong_latent_2006, fong_parallel_2005}, but we note that our conclusions remain
equally valid when this assumption is relaxed (Supporting Information, {\it
Effects of Nonoptimal Reference States} section). Table 1 summarizes the results for
all 52 single-gene knockouts considered in our study. Relative to the unmodified
strain in the suboptimal state following a gene knockout, the strain lacking 
the latent pathways exhibits equal or improved growth in 100\% of the cases
according to MOMA (Fig. \ref{fig2} {\it A}) and in 98\% of the cases according to
ROOM (Fig. \ref{fig2}{\it B}). Across all knockout perturbations, this
corresponds to an average change of $+8.5$\% and
$+1.2$\% of the optimal wild-type growth rate, respectively. With either
approach, a large fraction of mutants (50\% for MOMA,
77\% for ROOM) show a negligible difference in growth rate (within $\pm 1$\% of
the wild-type growth rate) when the latent pathways are disabled.
If only cases exhibiting significant
changes are considered, the removal of latent pathways consistently increases
the early postperturbation growth rate for all mutants, by an average of
$+16.9$\% for MOMA (Fig. \ref{fig2}{\it A}) and $+4.6$\% for ROOM (Fig.
\ref{fig2}{\it B}). Thus, for almost all knockouts, the strain lacking
latent pathways is predicted to suffer no competitive disadvantage compared to
the latent pathway-enabled strain. On the contrary, we predict that it more
often shows improved growth in the suboptimal regime shortly after the
perturbation.

\begin{figure}[t]
$\begin{array}{c}   
\includegraphics[angle=0,width=7.0cm]{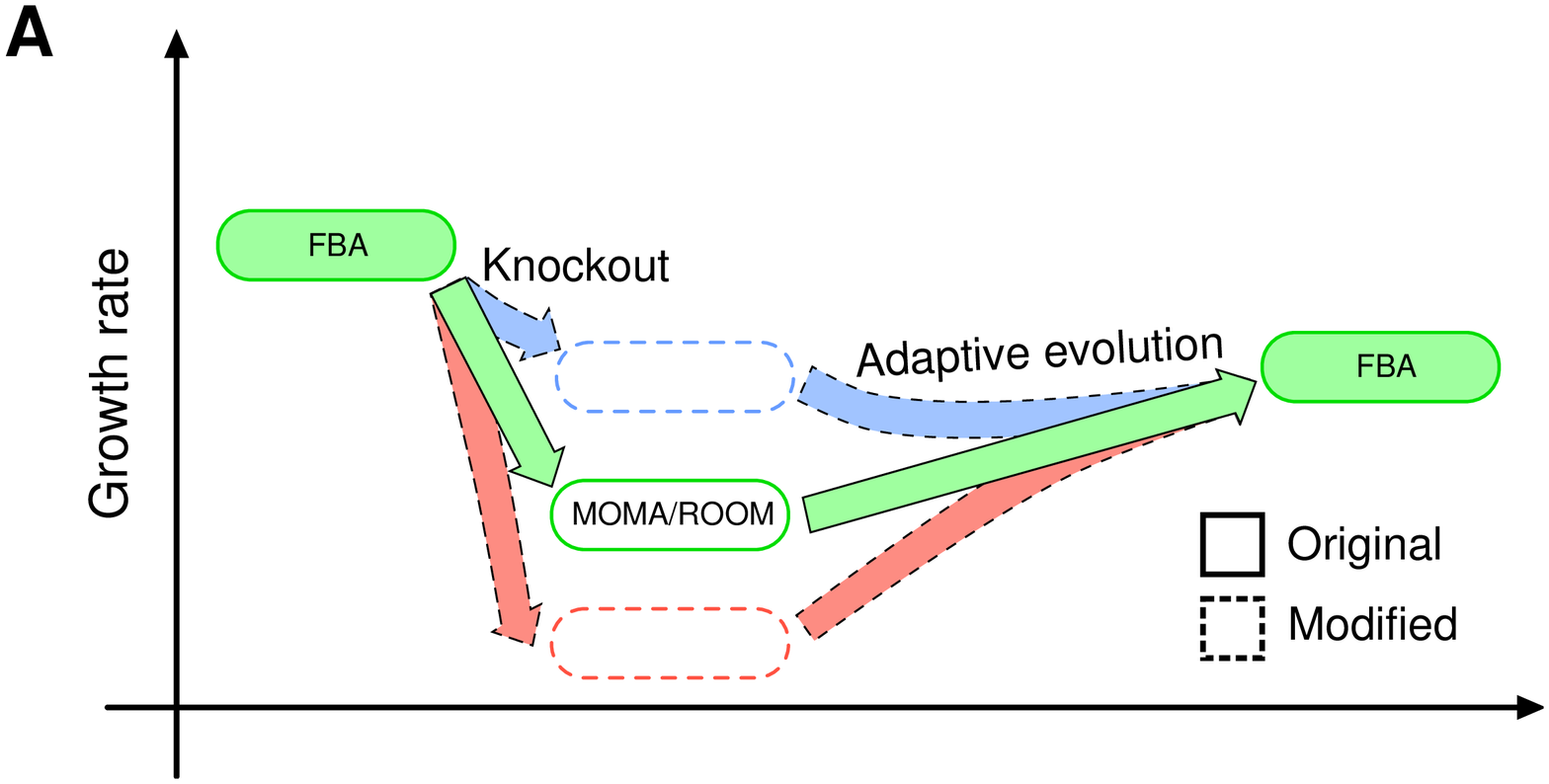} \\
\includegraphics[angle=0,width=7.0cm]{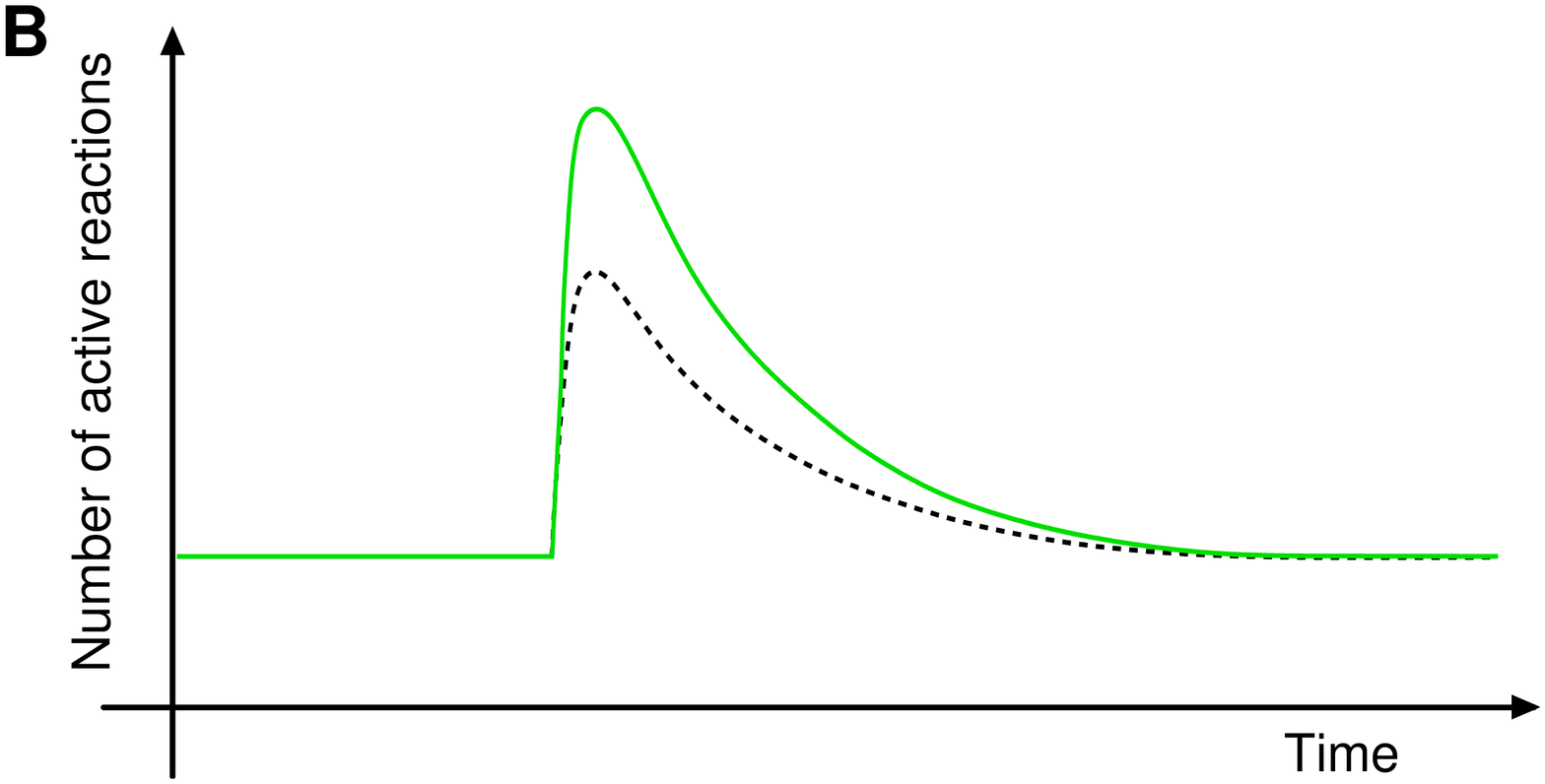}
\end{array}$
\caption{Hypothetical growth impact of latent pathways under a knockout
perturbation. ({\capitfont A}), ({\capitfont B}) The solid lines indicate the drop in
growth rate ({\capitfont A}) and the burst in latent reaction activation
({\capitfont B}) that follow a gene knockout. The dashed lines indicate the possible
behavior for a modified strain in which the original latent pathways have been
removed: the postperturbation growth rate may decrease (red), remain the same
(green) or increase (blue), and a smaller number of new latent pathways may be
created. If the postperturbation growth rate is nonzero, the mutant is viable,
and after a period of adaptive evolution it will converge to a new optimal
growth state. This postperturbation optimal growth state is identical for both
the modified and unmodified strain, and is characterized by a reduced number of
active reactions relative to the suboptimal states.}
\label{fig1}
\end{figure}

\begin{table*}[b!]
\caption{Summary of the predicted impact of latent pathways in {\tabtextifont E. coli} K-12 MG1655
under single-gene knockout perturbations.}
\begin{tabular*}{0.95\textwidth}{@{\extracolsep{\fill}} lccc }
& MOMA & ROOM & Random\\
\hline
Latent reactions for individual perturbations: & & \\
\;\; 
All knockout perturbations & +8.5\, (12.5)\%~ & ~+1.2\, (2.8)\%~ & ~+70.0\, (10.8)\% \\
\;\;
Significant differences\tablenote{By more than 1\% of the wild-type growth rate: 50\% (MOMA), 23\% (ROOM), and 100\% (random) of the perturbations.}
 & +16.9\, (13.2)\%~ & ~+4.6\, (4.5)\%~ & ~+70.0\, (10.8)\% \\
\;\; 
Number of reactions removed & 291\, (83) & 120\, (59) & 1,019\, (5)\\
\hline
Simultaneously nonessential latent reactions: & & \\    
\;\; 
All knockout perturbations  & +7.4\, (10.9)\%~ & ~+1.2\, (2.8)\%~  & \\
\;\; 
Significant differences\tablenote{By more than 1\% of the wild-type growth rate: 50\% (MOMA), and 19\% (ROOM) of the perturbations.} & +14.8\, (11.4)\%~ & ~+5.2\, (4.6)\%~ & \\
\;\; 
Number of reactions removed & 258\, (79) & 109\, (53) &  \\ 
\hline
\end{tabular*}
\label{table1}
\begin{flushleft}
\begin{tabular*}{0.95\textwidth}{@{\extracolsep{\fill}} l}
\begin{minipage}[t]{0.95\textwidth}%
Each column corresponds to the average (and standard deviation) of the difference 
in growth rate between the latent pathway-disabled and wild-type organisms for 52 
different single-gene knockouts. The differences are expressed as percentages of 
the optimal wild-type growth rate. For all cases, the average postperturbation growth rate is
higher for the strain without latent pathways.
\end{minipage}
\end{tabular*}
\end{flushleft}
\end{table*}

The set of transiently active (latent) reactions depends on the perturbation.
Even though we predict that in general, the removal of one of these 52 sets
increases growth under the corresponding knockout, the same removal may in
principle have an adverse effect under a different knockout. To address this
possibility, we first note that the sets of {\it simultaneously nonessential
latent reactions} remain sizeable: an average of 258 $\pm$ 79 for MOMA and 109
$\pm$ 53 for ROOM. For a given knockout perturbation, this set is defined as the
subset of the original latent reactions that are inactive in the optimal growth
state we consider for each of the other 51 knockout mutants. These reactions are
therefore dispensable for optimal growth, both in the wild type and all 52 single
gene knockout strains we consider, but are nonetheless transiently activated in
response to the given perturbation. We have tested the impact of disabling these
reduced sets of latent reactions under the corresponding knockouts (Materials
and Methods). As shown in Fig. \ref{fig2} {\it C} and {\it D}, the presence of these
simultaneously nonessential latent reactions has the same trend of inhibiting
growth adaptation as found for the full sets of transiently activated
reactions. 

The possibility that latent pathway activation enhances cells' viability
following a perturbation is a compelling hypothesis, as it would reveal
functions for genes that have thus far eluded high-throughput phenotype screens.
We note, however, that our analysis also predicts the transient activation of
pathways that do, in fact, have known phenotypes under different conditions. For
example, activation of the  glyoxylate shunt is known to  mitigate growth
defects of {\it E. coli} on glucose following phosphofructokinase mutations
\cite{vinopal_glyoxylate_1974}. Since we focus on single knockouts, the genes
affected by such mutations, {\it pfkA} and {\it pfkB}, are not among the
perturbations we consider. Nonetheless, out of the 52 unrelated knockout
perturbations in our study, our models show the transient activation of the
glyoxylate shunt in response to 25 of them according to MOMA and 7 according to
ROOM. The same phenomenon can be observed for reactions that are essential under
different environmental conditions but inactive in the aerobic glucose medium
employed in our simulations. Pyruvate formate lyase is required for anaerobic
growth in xylose medium according to experiments \cite{hasona_pfl_2004} and our
models, but is transiently active for 2 (MOMA) and 18 (ROOM) of the  52 genetic
perturbations in this study. This interesting effect---the nonspecific use of
pathways under an array of perturbations quite different from the conditions
under which they have an observed phenotype---indicates that the phenomenon of
latent pathway activation extends beyond the set of apparently nonessential
genes. 

\begin{figure*}[t]
	\begin{center}
		$\begin{array}{cc}
		\includegraphics[angle=0,width=7cm]{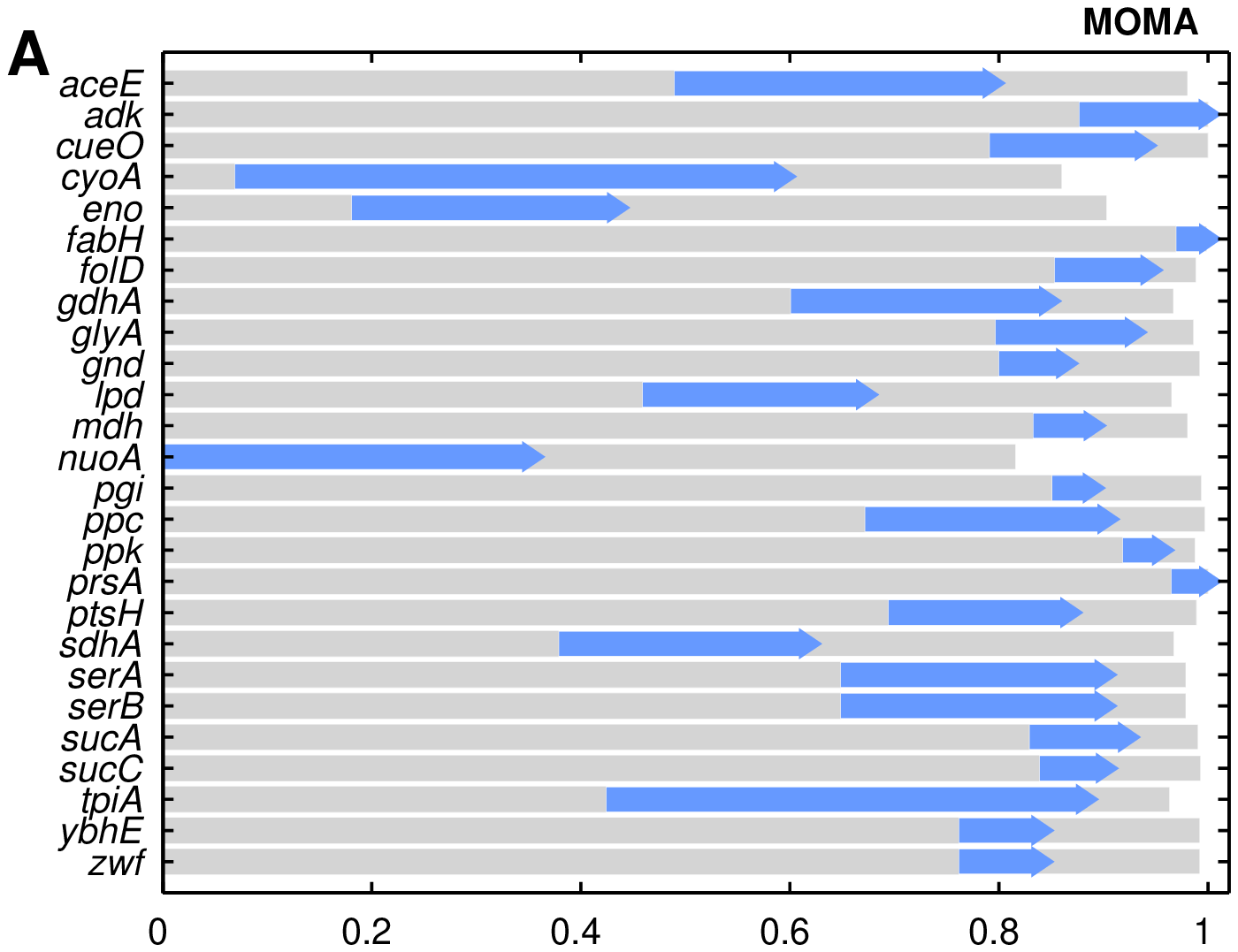} &
		\includegraphics[angle=0,width=7cm]{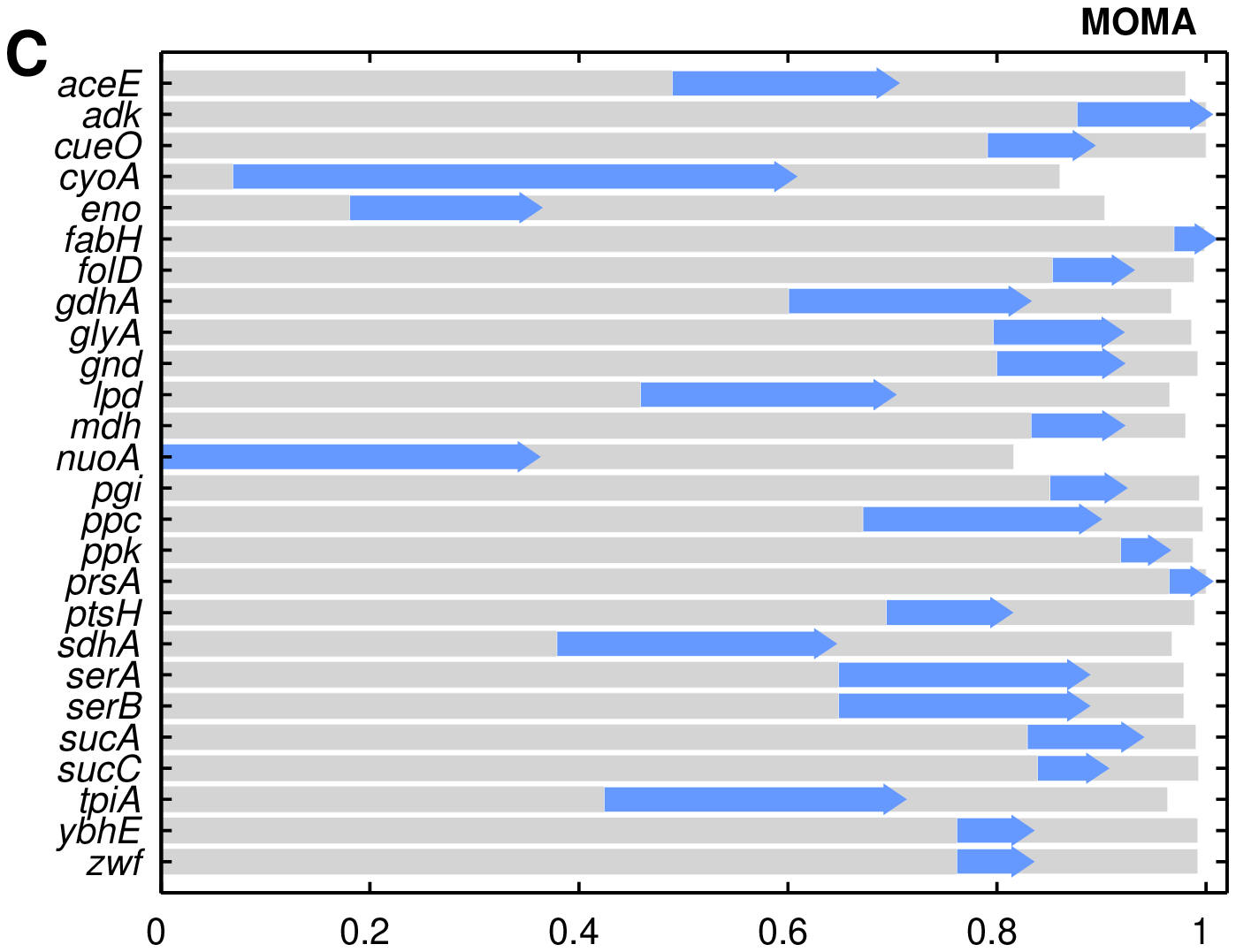} \\
		\includegraphics[angle=0,width=7cm]{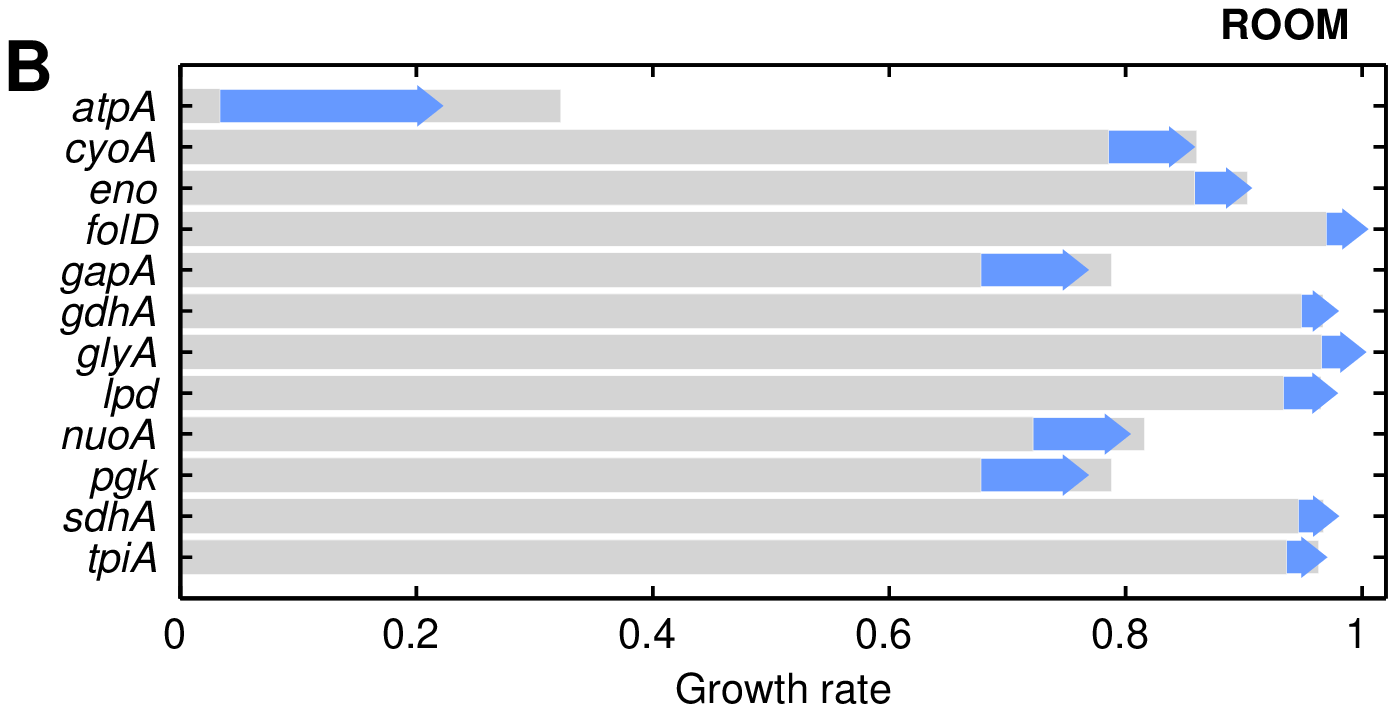} &
		\includegraphics[angle=0,width=7cm]{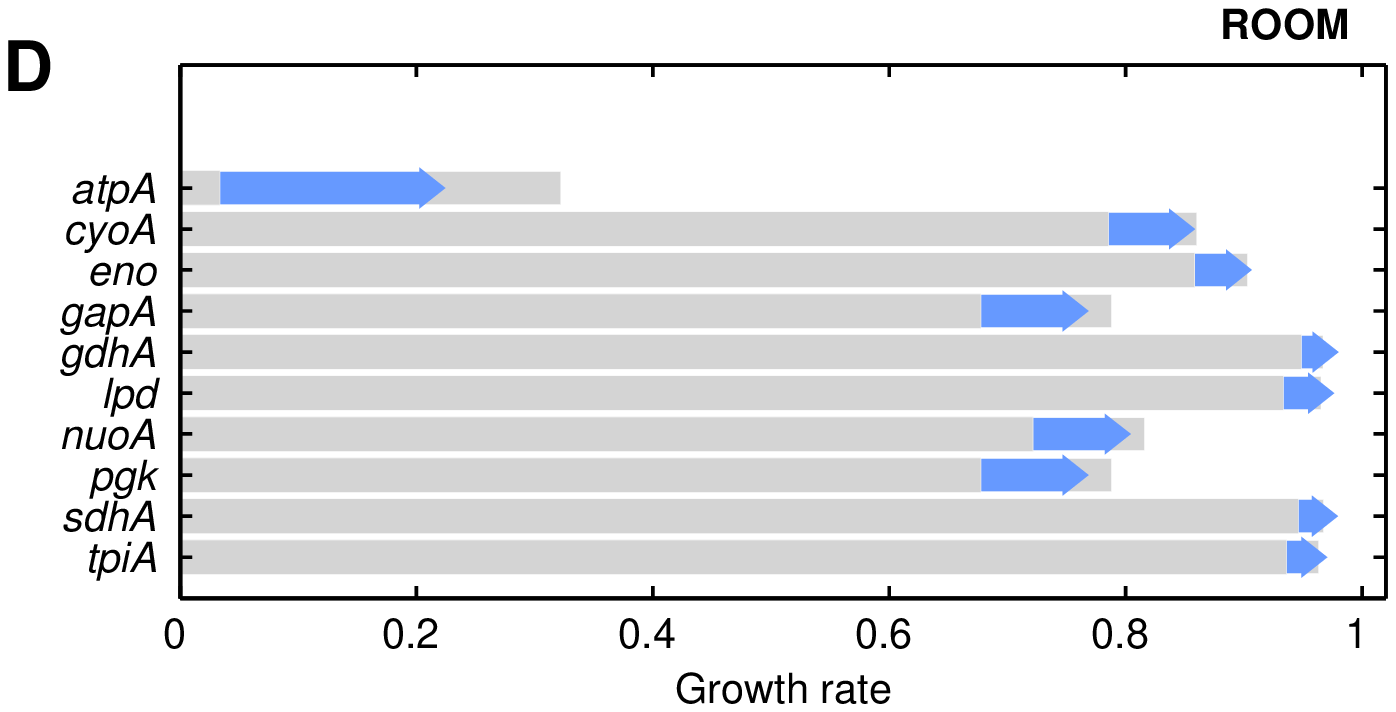}
		\end{array}$
	\end{center}
	\caption{Predicted adaptive impact of latent pathways in {\capitfont E. coli} under
	single-gene knockout perturbations. Each row indicates the difference in early
	postperturbation growth rate between the strains with and without latent
	pathways for the knockout perturbation indicated along the vertical axis, when:
	({\capitfont A} and {\capitfont B}) the modified organism lacks all reactions that are
	transiently active under the corresponding knockout perturbation; ({\capitfont C} and
	{\capitfont D}) the modified organism lacks the set of simultaneously nonessential
	latent reactions, which is the subset of such latent reactions that are not
	needed for growth in any of the other 51 single-gene knockout mutants. For both
	MOMA ({\capitfont A} and {\capitfont C}) and ROOM ({\capitfont B} and {\capitfont D}), only cases
	showing a significant change in growth rate ($>$ 1\% of the wild-type optimal
	growth rate) are shown. Arrows indicate the extent of the increase in growth
	rate when the latent pathways are removed. The shaded background indicates the 
	theoretical maximum growth rate for the mutant strain predicted by FBA. All
	growth rates are normalized by the optimal wild-type growth rate.  The
	statistics are summarized in the left and center columns of Table 1 for MOMA and
	ROOM, respectively. }
\label{fig2} 
\end{figure*}

\subsection*{Model-independent analysis.} 
\begin{figure*}[t]
$
\begin{array}{cc}
\includegraphics[angle=0,width=8.0cm]{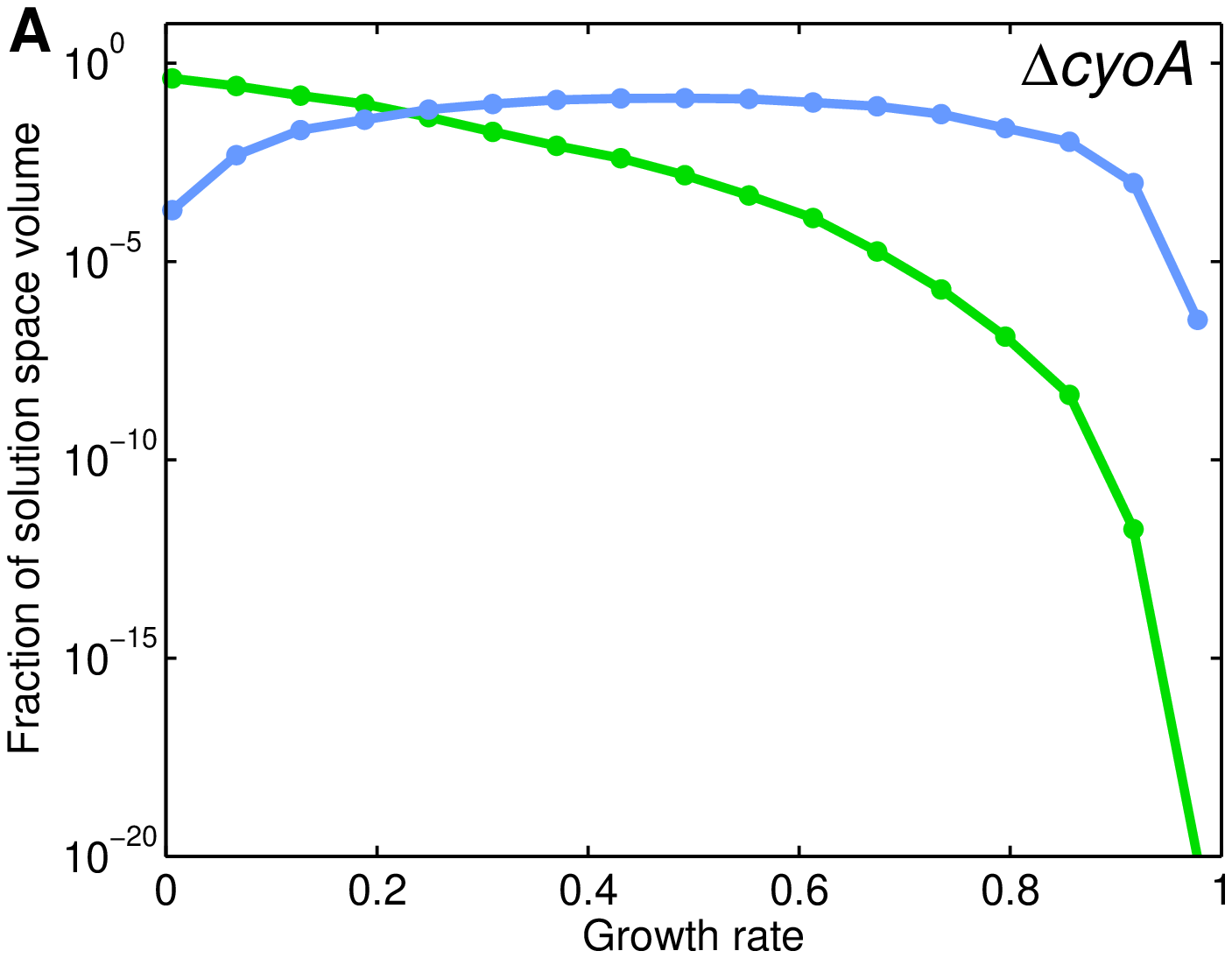}  &
\includegraphics[angle=0,width=8.0cm]{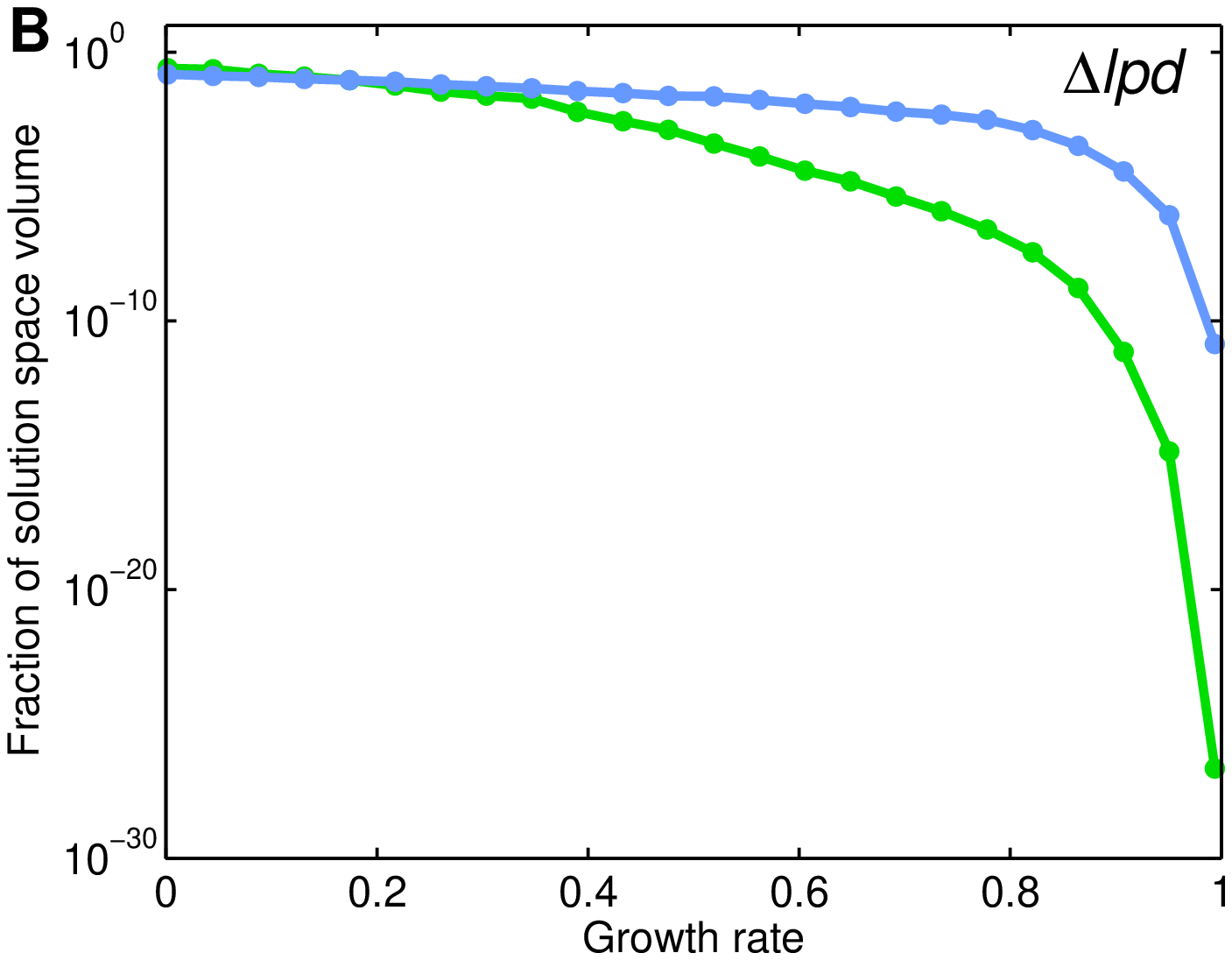}\hspace{0.1cm}
\end{array}$ \\
$\begin{array}{cc}
\includegraphics[angle=0,width=12.5cm]{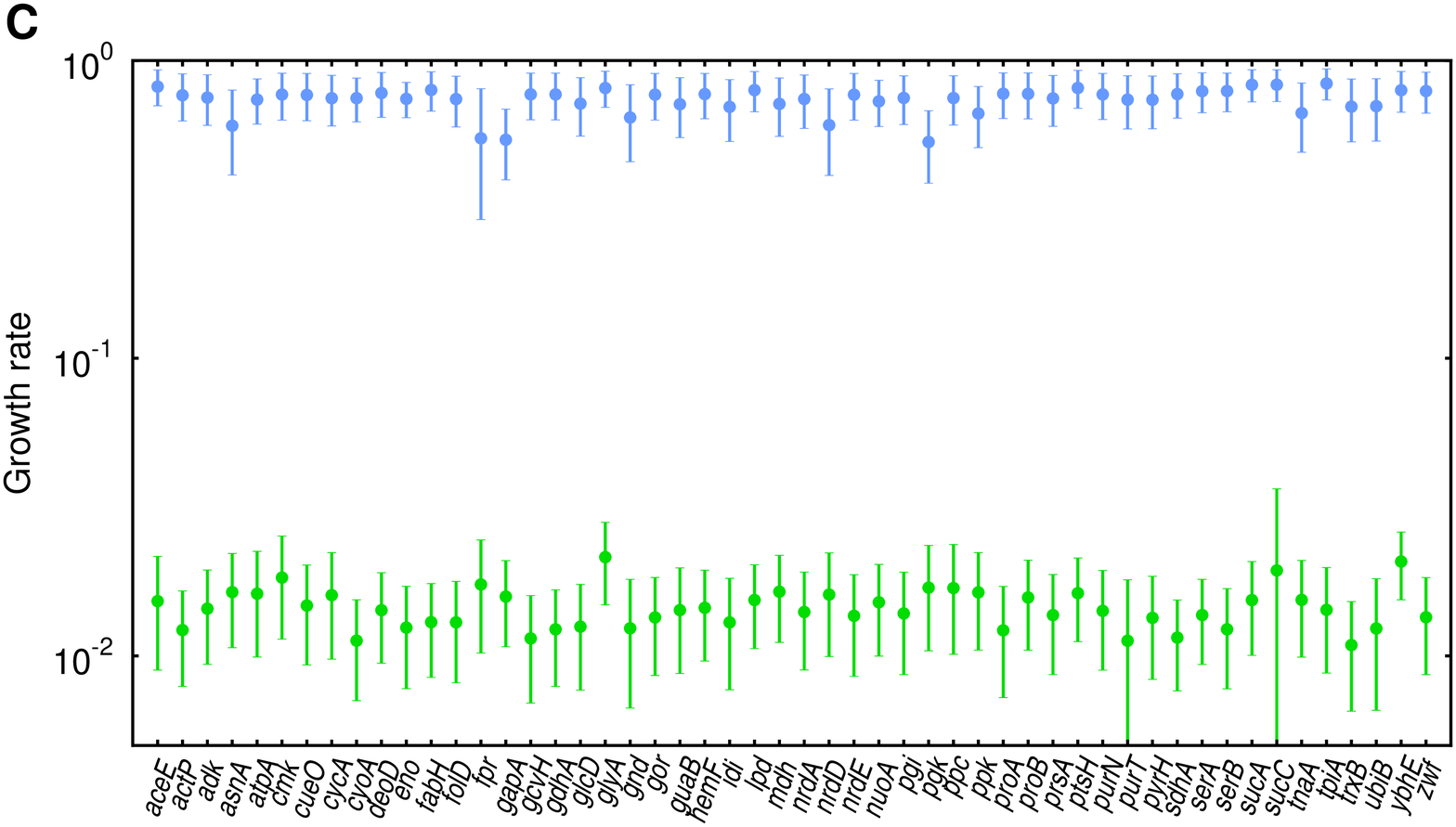} & 
\hspace{-1.0cm}\includegraphics[angle=0,width=4.5cm]{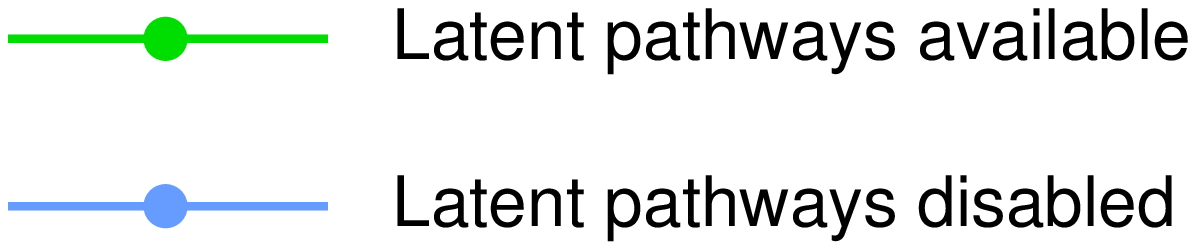}\hspace{0.1cm}
\end{array}$
\caption{Properties of the space of feasible metabolic states. ({\capitfont A} and
{\capitfont B}) Volume of the metabolic solution space as a function of the growth
rate for the {\capitfont cyoA}- ({\capitfont A}) and the {\capitfont lpd}- ({\capitfont B})  knockout
mutant. All values are normalized so that the area under each curve is 1.
Thus, each curve can be interpreted as a probability distribution of growth rate
within the corresponding metabolic space, either with or without latent
pathways. For increasing growth rate, the normalized volume decreases
systematically when the latent pathways are available (green lines), but the
curve becomes less steep and the volume may even peak at nonzero growth when latent pathways are disabled
(blue lines). These results were obtained for the MOMA-predicted response of the
network, and similar behavior is observed for the other single-gene knockout
mutants in our model. Disabling latent pathways therefore increases the relative
frequency of high-growth metabolic states at the  expense of low-growth states.
For numerical feasibility, this calculation was implemented using the central
metabolism of {\capitfont E. coli} ({\capitfont Materials and Methods}). ({\capitfont C})
Postperturbation growth rate if the network were to respond by moving to a
random position in the space of feasible states. The dots and error bars
correspond to the average and standard deviation for each of the knockout
mutants indicated at the bottom, when latent reactions are available (green) and
when latent pathways associated with the particular perturbation and random
response are disabled (blue). Interpreting a random response as a lower bound
for the likely response of the organisms, the systematically higher growth of
the latent pathway-disabled strains corroborates the conclusion that latent
pathways do not facilitate adaptation. These statistics are summarized in the
right column of Table 1. All growth rates are normalized by the FBA-predicted
rate of the knockout mutants.}
\label{fig3} 
\end{figure*}

\ The analysis above shows that the
availability of latent pathways inhibits growth in the short term after a
genetic perturbation. But how sensitive are these conclusions to the models we
used to simulate the response of the network? To provide model-independent
evidence, we have determined how the volume of the space of feasible
metabolic states ({\it Materials and Methods}) depends on the growth
rate. As shown in Fig. \ref{fig3} {\it A} and {\it B} (green lines) for the {\it cyoA} and
{\it lpd} knockout mutants, the volume systematically decreases as a function of
growth rate for the single-gene knockout mutants considered in our study,
indicating that the number of metabolic states available to the unevolved mutant
is much larger at lower growth rates. When the latent reactions
are disabled, however, the relative volume, and hence the relative number of
available metabolic states, increases for large growth rates (Fig.
\ref{fig3} {\it A} and {\it B}; blue lines).  Therefore, the principal effect of removing
latent pathways appears to be an increase in the relative frequency of
high-growth states due to the preferential elimination of low-growth states. It
should be noted, however, that a large number of high-growth states are also
disabled in this process due to the ``entanglement'' between latent pathways and
biomass-producing pathways that exist under the metabolic steady-state conditions of our
models (Supporting Information, {\it Elementary Mode Analysis} section). 

To appreciate the constraints imposed by this structure, imagine that the
organism responds to perturbations by moving to a random metabolic state in the
feasible space of fluxes. We simulated this hypothetical response using an
implementation of the hit-and-run sampling algorithm ({\it Materials and Methods}). As
shown in Fig. \ref{fig3}{\it C}, the postperturbation growth rate is nearly
{\it zero} for all mutants with latent pathways and close to the theoretical
{\it maximum} for all mutants without them. This random response is arguably a
lower bound for the actual response of organisms that have evolved to cope with
perturbations, but the conclusion is clear: unless we assume that organisms have
evolved to respond to perturbations in a highly specific manner, which appears
to be inconsistent with experiments \cite{stern_genome-wide_2007}, the
availability of latent pathways does not facilitate growth, and this prediction
is largely independent of the network response to perturbation. This holds true
in particular for MOMA and ROOM, which incorporate (in different ways) the main
flux rerouting features observed in the activation of latent pathways in {\it E.
coli} \cite{fong_latent_2006}.  

Further mechanistic insight comes from the recently identified synthetic
rescue interactions \cite{motter_predicting_2008}, in which the knockout of a
gene inhibits growth but, counterintuitively, the targeted concurrent knockout
of additional genes recovers the ability of the organisms to grow. The reactions
catalyzed by such rescue genes are predicted to be active in typical suboptimal
states and inactive in growth-maximizing states of the knockout mutant
\cite{nishikawa_spontaneous_2008,motter_predicting_2008}. Now, given the
observation above that the set of active reactions predicted by FBA is only
slightly modified by a gene knockout, it follows that most rescue genes are in
fact associated with latent pathways. This in turn explains why the latent
pathway-disabled strains show improved growth. Note that this argument cannot
be anticipated from intuition, because an enormous number of low-growth states (up
to several orders of magnitude larger than for near-optimal growth) may exist even
when latent pathways are disabled (Fig. \ref{fig3} {\it A} and {\it B}). Furthermore, even
in the extreme case when one disables all reactions that are inactive in
the optimal state of the knockout mutant, metabolic states with a very low
growth rate ($\approx$ 10\% or less of the wild-type optimum) exist in 47 out of the 52
mutants (Supporting Information, {\it Elementary Mode Analysis} section). This
threshold is significant since all but 5 unmodified knockout mutants exceed this
growth rate according to MOMA, and all but 1 according to ROOM. Thus, although our
model-independent analysis suggests that latent pathway activation inhibits
growth under a general choice of metabolic state after a perturbation,
this is not directly imposed by the geometry of the solution space. Rather,
the predicted growth benefit associated with latent pathway removal and
synthetic rescues is partly a reflection of the cells' adaptive response.

An extreme example of this rescue effect is provided by the {\it cyoA}-deficient
mutant, which is predicted by MOMA to drop to $<$ 10\% of the optimal wild-type growth
rate following the perturbation, but recovers to $\approx$ 60\% if the latent pathways are also
disabled. In addition, cases in which the single-gene knockout mutant operates
near the theoretical optimum but growth nonetheless improves upon the removal of
latent pathways, such as for the {\it folD} mutant, can be related to weaker forms of synthetic
rescues \cite{motter_predicting_2008}.

This surprising relation to synthetic rescues is particularly interesting when
we note that latent reactions define several pathways whose participation in
{\it E. coli}'s metabolism has been controversial or elusive. The
Entner-Doudoroff (ED) pathway, an alternative to glycolysis for glucose
catabolism, is inactive in wild-type {\it E. coli} according to in vivo
experiments in glucose \cite{flores_analysis_2002} but becomes transiently
active in mutants lacking phosphoglucose isomerase \cite{fong_latent_2006}. This
activation may serve to reduce NADPH accumulation accompanying increased flux
through the pentose phosphate pathway \cite{hua_responses_2003}. Both MOMA and
ROOM predict a small, transient flux through the ED pathway in response to the
knockout of {\it pgi}, the gene coding phosphoglucose isomerase. In triphosphate
isomerase-deficient strains, our model predicts the activation of the
normally inactive methylglyoxal bypass \cite{ferguson_methylglyoxal_1998}.
Experimentally, this pathway is observed to channel excess dihydroxyacetone phosphate (DHAP)
into pyruvate following glycolytic flux splitting into glyceraldehyde 3-phosphate
and DHAP after the knockout of the associated gene, {\it tpi} \cite{fong_latent_2006}. These findings emphasize the
importance of probing multiple gene knockouts or perturbations---previously
suggested in the context of synthetic lethality \cite{deutscher_multiple_2006},
synthetic rescues \cite{motter_predicting_2008,motter_improved_2010},
multi-target drug discovery \cite{hopkins_network_2008,motter_improved_2010},
and neutral mutations \cite{wagner_neutralism_2008}---as a means to determine
the puzzling role of transients. 

\section*{Discussion} \ 
Latent pathways and their associated genes are, by definition, dispensable for
optimal growth under the given conditions both before and after a perturbation.
Several explanations have been offered to justify the persistence of apparently
nonessential metabolic genes within the genome. Environmental robustness is the
most natural hypothesis, as it acknowledges the unpredictable, time-varying
conditions that confront single-cell organisms in natural environments. Yet,
in silico \cite{papp_metabolic_2004, freilich_2010} and experimental
\cite{harrison_plasticity_2007} studies of model organisms under various
environmental conditions  likely to occur in nature suggest that
condition-dependent robustness is inadequate to fully explain the dispensability
of metabolic genes. Part of the remaining redundancy may be attributed to the
varying efficiencies of redundant pathways under different  environments
\cite{wang_2009}, or selective pressure for increased metabolic capacity across
all conditions \cite{freilich_2010}. An alternative to environmental robustness
is genetic robustness, in which redundant pathways buffer against null mutations
\cite{barabasi_2004, gu_2003}. But regardless of which explanation might apply in
the context of latent pathways, the question that follows from our analysis is
not why these  dispensable pathways are present in the genome, but rather what
causes their {\it transient activation} in response to genetic perturbations.
Our results indicate that this activation confers no advantage in fitness as
measured by growth, and more often hinders growth in the short term following a
perturbation.

The dispensability of latent pathway activation predicted in this study can be
interpreted in three different ways, which are not necessarily mutually
exclusive. First, it is possible that temporary reaction activation does not
provide necessary intermediate states but is instead a byproduct of the
network's suboptimal response to perturbations. The fact that {\it E. coli} 
undergoes a period of suboptimal growth following genetic perturbations is
well-established  in both in silico and in vivo experiments
\cite{segre_analysis_2002,shlomi_regulatory/off_2005}. This interpretation is
thus supported by the recent observation that typical suboptimal metabolic
states have a much larger number of active reactions (more than $2.5$ times
larger for the conditions considered here) than typical states that optimize
growth \cite{nishikawa_spontaneous_2008}. 

A second interpretation is that cells activate a large number of reactions
as a global nonspecific response to perturbations that nevertheless creates a
library of possible metabolic states that can be subsequently fine-tuned by
adaptive evolution. This scenario cannot be rejected by existing experimental
results \cite{stern_genome-wide_2007} and is appealing as it allows for an
indirect, long-term benefit of temporary reaction activation even if, as
predicted here, this activation inhibits growth in the short term. Although
whole-cell regulation remains largely unexplained, there are known mechanisms
that could subsequently lead to optimal growth \cite{goyal_learning_2010,
pelosi_parallel_2006}. This interpretation suggests that fine-tuning of
whole-cell reaction activity is best achieved by down-regulating specific
over-expressed reactions rather than by the coordinated up-regulation of entire
under-expressed pathways, a principle that is widely appreciated in metabolic
engineering but that is yet to be demonstrated for natural systems. The recent
observation that {\it E. coli}'s response to perturbations is more stable from
the view point of metabolite concentration and protein or mRNA abundance
\cite{ishii_multiple_2007} than reaction flux  \cite{fong_latent_2006} may be
relevant for the validation of this interpretation. The objection to the second
interpretation is that it cannot explain (not even in the long term) the
availability of pathways that are simultaneously dispensable for all known
perturbations that have shaped the evolution of the organisms. This ``paradox of
latency" reinforces the need to study the {\it mechanisms} that govern latent
pathway activation.

In experimental studies of gene function in microbial organisms, growth is the
most often used indicator of fitness, owing to both its accessibility for
measurement and its intrinsic importance in determining viability
\cite{baba_construction_2006, gerdes_experimental_2003, giaever_functional_2002,
typas_high-throughput_2008}. Although the results of this study suggest that the
availability of latent pathways does not promote faster growth following a
perturbation, the conclusion that this negatively impacts adaptation is reached
by studying these pathways in the context of metabolism alone. Thus, a third
interpretation is that, in general, the cellular objective invoked in the
adaptive response is not growth. Instead, the activation of otherwise latent
metabolic pathways may accompany other cellular processes, either inside or
outside metabolism, that are initiated to ensure survival. For example, yeast
adopts changes in cell shape and internal pressure in response to osmotic shock,
a process that recruits the metabolism to accumulate specific metabolites
\cite{klipp_osmotic_2005}. Moreover, enzymes and metabolic cofactors involved
in transiently upregulated pathways may have regulatory or signaling roles in
addition to their metabolic function \cite{shi_2004}. This third possibility
allows for an advantageous, external function for pathways whose activation
appears disadvantageous when only the metabolism is considered.  In this way,
latent pathway activation is incorporated into a larger, more sophisticated
adaptive response. This explanation, however, seems inconsistent with our
observation of the apparently highly nonspecific activation of pathways in
response to different perturbations.

Regardless of which interpretation proves to be correct, this study leads to a
clear, experimentally testable prediction: that latent pathway activation does
not enhance, and in fact often inhibits, early postperturbation growth. While
we expect this behavior to be observed experimentally under a wide range of
conditions, regardless of the specific suboptimal response of the cell,
deviations from this behavior would also be highly informative since they would
uncover growth phenotypes not detected in previous steady-state experiments.
Varying environments can also generate strong \cite{ishii_multiple_2007} and
sometimes counterintuitive responses, such as the possibility of accelerating
the evolution of unevolved strains \cite{kashtan_varying_2007}. Therefore, it
may well be the case that in time-varying environments, which are beyond current
modeling capability, latent pathway activation will exhibit a different fitness
effect. 

\begin{materials}
\section{Network and perturbations} \ 
We used the \textit{i}AF1260 {\it E. coli} model \cite{feist_genome-scale_2007}, which
is the most up-to-date reconstruction of the metabolic network  of {\it E. coli}
 MG1655. The network consists of 2,082 unique reactions catalyzed by 1,260 genes
and involves 1,369 metabolites, as well as 299 exchange fluxes and the biomass
flux. We focused on the 52 single-gene knockout strains that are compatible with
growth but for which the original growth-maximizing flux state becomes
infeasible after the gene knockout. The volume calculation, which remains
challenging for the full network, was performed using a reduced network
consisting of 62 reactions, 101 genes, 49 metabolites, and 14 exchange fluxes
representing the central metabolism of {\it E. coli} \cite{reed_expanded_2003}. 

\section{Medium and constraints} \
The simulated medium consisted of limited amount of glucose (8 mmol/g DW-h) and
oxygen (18.5 mmol/g DW-h), and unlimited amount of carbon dioxide, iron (II),
protons, water, potassium, sodium, ammonia, phosphate, and sulfate. Irreversible
reactions are constrained to have nonnegative fluxes. The flux through the ATP
maintenance reaction was set to 8.39 mmol/g DW-h. A total of 1,432 reactions in
the \textit{i}AF1260 model, including 61 reactions in the subnetwork of the central
metabolism, can be active under these medium conditions. An analysis of the
effect of regulatory limitations \cite{covert_regulatory_2004} that may
constrain this reaction activity in vivo is considered in the Supporting
Information ({\it Regulatory Constraints} section). There it is shown that our results
remain valid under the additional constraints imposed by these limitations.

\section{Feasible metabolic states} \
The state of the metabolic network is represented by a vector of all reaction
fluxes $\boldnu = (\nu_j)$. Since the time scale at which the network responds
to perturbations is much shorter than the characteristic time for adaptive
evolution, we focused on steady-state flux
distributions both before and after perturbation. Steady-state fluxes are
solutions of the mass-balance equation $\boldS \cdot \boldnu = 0$, where $\boldS
= (S_{ij})$ is the matrix of stoichiometric coefficients, under the constraint
imposed by the medium, reaction irreversibility, ATP maintenance requirements,
and possibly gene knockouts. A knockout of the enzyme-coding genes associated
with reaction $j$ is implemented through the additional constraint $\nu_j = 0$.
The solution space is the set of all such steady-state flux vectors and it has
the form of a convex polytope.  We refer to the individual solutions in this
space as {\it feasible metabolic states}.

\section{Objective functions} \
FBA \cite{edwards_escherichia_2000} identifies a growth-maximizing state within
the space of feasible metabolic states by maximizing the flux $\nu_b$ through a
biomass reaction that drains biomass precursors. With respect to this original
state, MOMA \cite{segre_analysis_2002} and ROOM
\cite{shlomi_regulatory/off_2005} identify feasible states that minimize the
distance in the space of fluxes and the number of significant flux changes,
respectively (Supporting Information, sections 2-3).  Our implementations of
FBA, MOMA, ROOM, and the hit-and-run algorithm used a commercial  optimization
package (ILOG CPLEX, version 11.0, www.ilog.com). For all FBA results, we have
used the growth-maximizing states provided by the simplex algorithm. More
information about the computational methods is provided in Supporting
Information, where it is also shown that our results do not depend sensitively
on the assumption of optimal growth in the reference states either before or
after the perturbation ({\it Effects of Nonoptimal Reference States}), nor on
particular choices for the growth-maximizing states used throughout the paper
({\it Sensitivity to Alternate Optima}). 

\section{Biomass flux and growth rates} \
The in silico model predicts the biomass flux, but for exponential growth
the result can be expressed in terms of a normalized growth rate $\bar{\kappa}$.
This follows from the observation that biomass production is governed by 
$\frac{1}{N} \frac{dN}{dt}=\kappa$, where $\kappa$ is the growth rate, $N$
measures the population size, and $\frac{1}{N} \frac{dN}{dt}$ is proportional to the biomass flux $\nu_b$.
Therefore, when normalized with respect to the wild-type or theoretical maximum,
the normalized biomass flux $\nu_b/\nu_{b,0}$ equals the corresponding
normalized growth rate $\kappa/\kappa_0$ used throughout the paper. 

\section{Identification of latent pathways} \
We define the latent reactions (pathways) associated with the knockout of gene
$A$ to be the set $L_A$ of all reactions predicted to be transiently active in
the unevolved mutant shortly after the perturbation (according to MOMA, ROOM, or
the hit-and-run algorithm), but inactive in both the optimal wild-type and
mutant strains (according to FBA). This set is therefore nonessential for
optimal growth under the knockout of $A$, although in principle some of these
reactions may be necessary for growth under the knockout of a different gene,
$B$. To account for this, we tested the impact of a set $L_A'$ of latent
reactions that are simultaneously nonessential in the optimal growth states of
the other 51 single-gene knockout mutants that we consider. 

\section{Volume calculation} \
The exact calculation of the volume of the (high-dimensional) solution space is
computationally intractable and, because this set is very skewed
\cite{bianconi_2008}, even  approximate calculations are computationally
demanding. To determine the volume of the solution space as a function of growth
rate, we used an approximate inference algorithm based on Belief Propagation
\cite{braunstein_estimatingsize_2008}. In this approach, the convex polytope
representing the constrained flux space is  tiled with hypercubes of size
$\varepsilon$. We then use a message-passing algorithm to approximate the
probability distribution $P({\boldnu})$ over the discretized space. Using this,
we define the associated entropy $S = -\sum P({\boldnu})\log_{10}P({\boldnu})$,
which counts the log of the number of $\varepsilon$-hypercubes that overlap the
space of feasible states. The total volume covered by these cubes is then $V =
10^{S}\varepsilon^n$, where $n$ is the dimension of the space. If $m$ is the
number of linearly independent mass balance constraints and $f$ is the number of
available fluxes after a given knockout, then the dimension of the unmodified
metabolic space is $n = f - m$ while the solution space for the modified
organism has dimension $n = f - m - l$, where $l$ is the number of independent latent
reactions removed. We used a granularity of $\varepsilon = \frac{1}{64}$ for all
calculations presented in the paper.

\section{Hit-and-run algorithm} \
To randomly sample the metabolic solution space, we implemented a hit-and-run algorithm
\cite{smith_efficient_1984} with artificial centering
\cite{kaufman_direction_1998}, which is a quickly converging sampler for
high-dimensional convex sets. The algorithm is  based on selecting a randomly
oriented line $l$ passing through the current sample point, and then selecting
the point for the next iteration from a uniform distribution along $l$.
For efficient sampling, we employed artificial centering
\cite{kaufman_direction_1998}, where the orientation of the line $l$ is obtained
from the direction defined by the current sample point and the center (mean) of a
subset $C$ of already-sampled points. The subset $C$ was initially created by
taking 10,000 warm-up points on the boundary of the space, and was updated
recursively by replacing a randomly selected point of $C$ with the currently
sampled point. In all calculations presented in the paper, we generated a set of
5 x 10\superscript{6} points to sample the solution space. All calculations were
performed with the COBRA Toolbox \cite{becker_2007}. \end{materials}

\begin{acknowledgments}
This work was supported by the National Science Foundation under Grant DMS-0709212, 
the National Cancer Institute under Grant
1U54CA143869-01, and a Sloan Research Fellowship to A.E.M.\\
\end{acknowledgments}
\subsection{Author contributions}
S.P.C. and A.E.M. designed research; S.P.C. and J.S.L. performed research;
S.P.C. and J.S.L. analyzed data; and S.P.C. and A.E.M. wrote the paper.
\subsection{Conflict of interest statement}
The authors declare no conflict of interest.

\end{article}

\begin{article}
\renewcommand{\theequation}{S\arabic{equation}}
{\noindent\huge\baselineskip= 24pt =\frutigerboldcondensed at 22pt \titlefont Supporting Information \par}
\medskip
\noindent
A metabolic network of $m$ metabolites and $n$ reactions is represented by a $m
\times n$ matrix $\boldS = (S_{ij})$, where $S_{ij}$ is the stoichiometric
coefficient of metabolite \textit{i} in reaction $j$. The state of the metabolic network
is represented by a vector of reaction fluxes $\boldnu = (\nu_j)$, where $\nu_j$ is the flux through
reaction $j$.  In our analysis, the steady-state solutions of the system are
determined by mass balance constraints,
\begin{equation}
\boldS \cdot \boldnu = 0,
\label{massbalance}
\end{equation}
and additional constraints that limit the range of the individual fluxes,
\begin{equation}
\boldnu_{\min} \leq \boldnu \leq \boldnu_{\max}, 
\label{thermo}
\end{equation}
where the inequalities in this notation are assumed to apply to each component
individually. The bounds on individual fluxes are determined by substrate
availability in the given medium, the ATP maintenance requirement, and
thermodynamic constraints that limit the reversibility of the corresponding
reaction. The knockout of the genes associated with enzymes catalyzing reaction
$j$ corresponds to the additional constraint $\nu_j = 0$. The exchange fluxes
and the biomass flux are excluded from our implementation of the MOMA and ROOM
objective functions. \\

\noindent
{\subsectionfont Flux Balance Analysis (FBA).} 
FBA \cite{edwards_escherichia_2000_si} is used to identify flux vectors $\boldnu$
satisfying (\ref{massbalance}) and (\ref{thermo}) that maximize biomass
production, which is represented by an additional reaction $b$ that drains biomass
components. The problem is implemented as a linear program:
\begin{eqnarray}
\textrm{max} & & \nu_b \\
\textrm{s.t.} & & \boldS \cdot \boldnu = 0 \nonumber \\
& & \boldnu_{\min} \leq \boldnu \leq \boldnu_{\max}. \nonumber
\end{eqnarray}
In general, the FBA solution is not unique, and the results in this paper were
obtained by selecting the optimal solution provided by the simplex algorithm.
For a discussion of the effects of choosing alternate FBA solutions, see
Sensitivity to Alternate Optima below.
\\

\noindent
{\subsectionfont Minimization of Metabolic Adjustment (MOMA).}
MOMA \cite{segre_analysis_2002_si} selects a suboptimal growth state $\boldnu$ by
minimizing the Euclidean distance in flux space from the wild-type optimal
growth state, $\boldw$. This is implemented as a quadratic programming problem: 
\begin{eqnarray}
\textrm{min} & & (\boldnu-\boldw)^T \cdot (\boldnu-\boldw) \\
\textrm{s.t.} & & \boldS \cdot \boldnu = 0 \nonumber \\
& & \boldnu_{\min} \leq \boldnu \leq \boldnu_{\max} \nonumber \\
& & \nu_j = 0, \; j \in A, \nonumber
\end{eqnarray}
where $A$ is the set of indices corresponding to reactions deactivated by gene
knockouts.\\

\noindent
{\subsectionfont Regulatory On/Off Minimization (ROOM).}
Unlike MOMA, which favors a potentially large number of small-magnitude flux
changes, ROOM \cite{shlomi_regulatory/off_2005_si} chooses a suboptimal growth
solution $\boldnu$  with a minimal number of ``significant" flux changes from
the original state.  This is usually implemented as a mixed-integer programming
problem  ({\it integer ROOM}):
\begin{eqnarray}
\textrm{min} & & \sum_{j=1}^n y_j \\
\textrm{s.t.} & & \boldS \cdot \boldnu = 0 \nonumber \\
& & \boldnu_{\min} \leq \boldnu \leq \boldnu_{\max} \nonumber \\
& & \nu_k = 0, \; k \in A \nonumber \\
& \textrm{for j}=1,...,n & \nonumber \\
& & \nu_j - y_j(\nu_{\max,j}-w_j^u) \leq w_j^u \nonumber \\
& & \nu_j - y_j(\nu_{\min,j}-w_j^l) \geq w_j^l \nonumber \\
& & w_j^u = w_j + \delta |w_j| + \epsilon \nonumber \\
& & w_j^l = w_j - \delta |w_j| - \epsilon \nonumber \\
& & y_j \in \lbrace 0,1 \rbrace, \nonumber
\end{eqnarray}
where $\boldw$ and $A$ are as for MOMA, and $\delta$ and $\epsilon$ express
tolerances for relative and absolute change from the original state,
respectively.

In our numerical experiments we chose to use a linear programming variant of the
method ({\it linear ROOM}) obtained by allowing the above binary constraints to
be continuous in the interval $0 \leq y_j \leq 1$ and setting $\delta = \epsilon
= 0$. Linear ROOM is biologically well-motivated, given that gene activity is
best described by a continuous variable, and has the advantage of being
computationally inexpensive for all mutants. Table S1 shows a comparison of the
growth impacts predicted by integer ROOM and the linear variant upon latent
pathway removal.  For most mutants, the two methods show similar change in
growth rate (increase, decrease, or an insignificant change) when the latent
pathways are removed, even though the exact growth rate predictions may differ.

For the integer variant, we followed Shlomi {\it et al.}
\cite{shlomi_regulatory/off_2005_si} in choosing the values $\delta = 0.03$ and
$\epsilon = 0.001$, which yielded reasonable running times for most knockout
strains. Even for larger values of these tolerances values there are a handful
of cases for which no optimal solution is found  in any reasonable amount of
time.  For these cases, in the comparison of Table S1, we take the best solution
found after one hour of computation on a 3.4 GHz CPU. The integer ROOM solutions
were further constrained to minimize the aggregate flux change from the
wild-type optimal state, $\sum |\nu_j-w_j|$. \\

\noindent
{\subsectionfont Sensitivity to Alternate Optima.}
In general, the optimal flux distribution given by FBA is not unique, neither
before nor after a given gene knockout \cite{mahadevan_effects_2003_si,
reed_genome-scale_2004_si}. The implications of this nonuniqueness must be
considered for two reasons. First, the set of transiently active reactions is
defined with respect to the exact optimal flux distribution before and after the
knockout perturbation, as well as the suboptimal flux distribution after the
perturbation (see Effects of Nonoptimal Reference States below for an analysis
of the effects of choosing suboptimal reference states in this definition of latent
pathways). Second, the postperturbation flux distribution is itself dependent
on a particular choice for the original optimal state, since both MOMA and ROOM
operate by minimizing some distance to a reference flux state.

To test the sensitivity of our results to a particular choice of FBA solutions,
we repeated our simulations for numerous combinations of wild-type and optimal
mutant flux distributions. We sampled the available FBA-predicted states by
fixing the corresponding growth rate (either wild type or optimal mutant) and
maximizing or minimizing each of the reaction fluxes allowed to vary under this
additional constraint. After choosing a particular pair of optimal flux
distributions in this way, we determined the corresponding set of latent
pathways according to MOMA and ROOM, and the associated suboptimal growth rates
before and after the removal of these latent pathways. Figure \ref{figs1} shows
the results of this analysis  for the {\it ppk}- and {\it tpiA}-knockout
perturbations. The distributions of growth rate predictions for the strains with
(green) and without (blue) latent pathways do not overlap. A less extensive
sampling for each of the other 50 knockout mutants considered in  our study
shows similar behavior. Altogether, these results suggest that our predictions
about the growth effect of latent pathway availability are robust with respect
to alternate optimal flux distributions. \\

\noindent
{\subsectionfont Regulatory Constraints.}
All results for the full {\it E. coli} \textit{i}AF1260 model presented in the main
text were computed under the uptake and steady-state constraints listed in the
{\it Materials and Methods} section. These constraints do not take into account
regulatory effects, which may limit the set of genes that can be transcribed
under the nutrient conditions we consider. Regulatory effects are thus expected
to constrain the set of steady-state metabolic states predicted by FBA, MOMA,
and ROOM that can actually be realized in vivo
\cite{covert_regulatory_2004_si}. To address the possible impact of regulatory
constraints on our predictions, we have repeated all calculations in the paper
for a modified version of the \textit{i}AF1260 model, following Feist {\it et al.}
\cite{feist_genome-scale_2007_si} in disabling a set of 152 reactions beforehand.
There is evidence that, due to
regulatory constraints, these reactions are inactive in the aerobic glucose conditions we
simulate. Under these additional constraints, there are 46 single-gene knockouts
that change the original flux distribution but nonetheless allow nonzero growth
in the resulting knockout mutants according to FBA, which is the same criterion we used to select
knockout perturbations in the main text for the unmodified \textit{i}AF1260 model.

Table S2 summarizes the predicted growth impact of disabling latent pathways in
the modified \textit{i}AF1260 model. Regardless of the approach used to simulate the
response (MOMA, ROOM, or hit-and-run sampling), all 46 knockout mutants in this
model show nearly equal (within $\pm 1$\% of the wild type) or improved growth
in the short term following the perturbation when the latent pathways are
disabled. Moreover, the average growth increases and the numbers of transiently
activated reactions are comparable to those presented in Table 1 of the main text for the
unmodified \textit{i}AF1260 model. Our results are therefore not dependent on having
the full complement of metabolic reactions in the network available for
activation in the organisms' initial response to perturbations. Indeed, the
predicted adverse growth effect of latent pathway activation is expected to hold
even under additional constraints that may reflect other limitations of the
metabolic response in vivo. \\

\noindent
{\subsectionfont Effects of Nonoptimal Reference States.}
The phenomenological models used in the main text predict a suboptimal metabolic
state by minimizing the value of a distance metric with respect to the
preperturbation reference state, which is assumed to be growth-maximizing as
predicted by FBA. The postperturbation reference state was also assumed to be
growth-maximizing after adaptive evolution. These simplifications overlook the
possibility that in many natural environments the metabolic state may be nonoptimal
even before and long after an external perturbation. This situation could arise,
for example, under time-varying conditions, where environmental changes prevent the
organisms from approaching optimal growth states. In this case, some of the
pathways we have classified as latent may fail to qualify in vivo since
they carry nonzero flux in one or both of the reference states.

Geometrically, this scenario corresponds to the appropriate (nonoptimal) reference states
lying in the interior of the feasible metabolic solution space, where many more
reactions are active, as opposed to the boundary of the space, where reaction
activity is limited by irreversibility constraints
\cite{nishikawa_spontaneous_2008_si}. This situation can be accommodated with a variant of our modeling approach.
We have systematically tested the growth impact of reaction {\it upregulation} relative to nonoptimal reference
states, which is the natural generalization of the latent reaction {\it activation}
considered in our original analysis. Given two optimal states $\boldnu_1$ (wild
type) and $\boldnu_2$ (evolved mutant) and their associated biomass fluxes,
$\nu_{b,1}^{opt}$ and $\nu_{b,2}^{opt}$, we now limit the biomass flux to some
fraction of these optimal values. This is imposed within our in silico
models by the additional constraints $\nu_{b,1} = \lambda \nu_{b,1}^{opt}$ and
$\nu_{b,2} = \lambda \nu_{b,2}^{opt}$, where $0 < \lambda < 1$. The
nonoptimal reference states are then defined by replacing the optimal states
before and long after a knockout with the closest feasible metabolic states that
satisfy these growth constraints. We performed this analysis for $\lambda = 0.4$ and $0.7$.

With respect to the new choices of nonoptimal reference states, the short term
response of the metabolic network to single-gene knockouts still exhibits a
transient burst of reaction activity. According to MOMA, the average and
standard deviation of the number of fluxes with larger magnitude than in both
reference states is 260 $\pm$ 83 ($\lambda = 0.4$) and 263 $\pm$ 82 ($\lambda =
0.7$). These numbers are comparable to the $291 \pm 83$ reactions that are
transiently activated by MOMA with respect to optimal reference states (main
text, Table 1). Figure \ref{figs2} shows the MOMA-predicted difference in growth
rate when these  transiently upregulated fluxes are constrained to not exceed
the reference states in magnitude. In all cases, downregulation of
the transiently upregulated pathways is predicted to improve growth in the short
term following a knockout perturbation, by an average of 6.0\% ($\lambda = 0.4$)
and 10.4\% ($\lambda = 0.7$) of the optimal wild-type growth rate. This analysis
is therefore in agreement with the prediction presented in the main text for
optimal reference states, namely, that the transient activation (or otherwise
upregulation) of latent pathways generally inhibits growth in the short term
following a perturbation. \\

\noindent
{\subsectionfont Elementary Mode Analysis.}
The results of Fig. 3 (main text) suggest that the primary effect of latent
pathway removal is to favor the availability of high-growth metabolic states by
preferentially eliminating low-growth states following a genetic perturbation.
This is accomplished by eliminating reactions that are silent in the optimal
states we consider before and after the knockout, thereby increasing the likelihood that the
initial metabolic response will activate pathways associated with higher growth. It should
be noted, however, that a given reaction can in general be active 
in many metabolic states, spanning both low- and high-growth phenotypes. 
It is therefore likely that many high-growth states will by eliminated by disabling latent pathways as well.

To systematically examine this connection between high-growth states and latent
pathways, we have used Elementary Mode (EM) Analysis, which is an approach for
analyzing metabolic networks in terms of interconnected sets of reactions
\cite{trinh_elementary_2009_si}. An EM is defined as a 
unique set of active reactions in the network (represented by a vector $\boldnu$ in flux space) 
that {\it i)} satisfies the
steady-state constraints $S \cdot \boldnu = 0$, {\it ii)} obeys all
reversibility constraints (negative entries in $\boldnu$ must correspond to
reversible reactions), and {\it iii)} is minimal in the sense that no reaction
may be removed from the set while still satisfying {\it (i)} and {\it (ii)}
\cite{schuster_elementary_1999_si}. Any steady-state flux distribution can be
represented as a linear combination of EMs with nonnegative coefficients.

The number of EMs of grows combinatorially with the size of the metabolic
network, making their calculation computationally infeasible for the full {\it
E. coli} \textit{i}AF1260 model. Therefore, we focused on the subnetwork comprising
{\it E. coli's} central metabolism, which was also used for the volume
calculation (main text, {\it Model-Independent Analysis}). Using the program METATOOL
\cite{metatool_2006_si}, we obtained the full set of 18,656 EMs available on
glucose. We have classified each mode as ``biomass-producing'' or
``nonbiomass-producing'', based on whether it has a positive or zero entry
corresponding to the (irreversible) biomass flux, respectively. It follows that
a linear combination of EMs representing a general zero or low-growth metabolic
state will be composed primarily of non biomass-producing EMs, with only a small
aggregate contribution from the EMs that produce biomass. Figure
\ref{elementary-mode-analysis} shows the effect of latent pathway removal on
these two types of EMs for the {\it cyoA} and {\it lpd} mutants. A significant
fraction of the original 18,656 EMs are eliminated by the knockout perturbation
(namely, those that involve a disabled reaction). Further EMs will be eliminated
when the latent pathways are disabled. As expected from the solution space
volume calculation and hit-and-run analysis in the main text ({\it Model-Independent
Analysis} section), the effect of latent pathway removal is to skew the
distribution of EMs toward those that produce biomass (Fig.
\ref{elementary-mode-analysis}{\it A}). But, surprisingly, a large number of
biomass-producing modes are sacrificed as well, and in fact comprise the {\it
majority} of EMs disabled in this process (Fig.
\ref{elementary-mode-analysis} {\it A} and {\it B}). 

Additional insight comes from analyzing the growth capabilities of the
metabolism when one eliminates {\it every} reaction that is silent in the
corresponding FBA-predicted optimal state of the knockout mutant. This
corresponds to an average of 1,967 $\pm$ 6 reactions across all 52 knockout
mutants, roughly 7 and 16 times larger than the number of latent pathways
removed for MOMA and ROOM, respectively (main text, Table 1). Figure
\ref{extreme-removal} shows the range of growth rates that can be realized by
each mutant in this scenario. For the majority of the mutants, 
metabolic states with very low growth
($<$ 10\% of the optimal wild-type growth rate) exist, even when this
large set of optimally silent reactions is disabled. This is particularly
significant given that, with few exceptions (5 according to MOMA, 1 according to
ROOM), the unadapted states of the 52 tested knockout mutants exceed this
growth threshold in our models when the latent pathways are enabled.

Taken together, these effects confirm that latent pathways cannot be considered
in isolation from biomass production, particularly in optimal growth states. \\

\noindent
{\subsectionfont Comparison with \textit{i}JR904 {\it E. coli} Model.}
We have repeated our calculations for the extensively curated \textit{i}JR904
reconstruction of the {\it E. coli}  metabolic network
\cite{reed_expanded_2003_si},  which has been previously analyzed in great detail
in connection with synthetic rescue interactions. This network consists of 931
reactions, 904 enzyme-coding genes, 618 metabolites, 143 exchange
fluxes, and the biomass flux.
Within this model, there are 36 single-gene knockout strains that are compatible with growth 
but for which the original growth-maximizing metabolic state becomes infeasible after the knockout.
For this set of genes, the average and standard deviation changes in growth rate are
+12.0 (15.8)\% (MOMA) and +1.1 (3.1)\% (ROOM) and +65.6 (11.4)\% (random) for
the removal of the latent reactions associated with the individual knockout
perturbations. The results are therefore consistent with those presented in
Table 1 for the \textit{i}AF1260 model (main text). \\

\end{article}

\renewcommand{\thefigure}{S\arabic{figure}}
\renewcommand{\thetable}{S\arabic{table}}

\pagebreak
\setcounter{figure}{0}
\begin{figure*}[!ht]
\begin{centering}
$\begin{array}{cc}
\includegraphics[angle=0,width=8.0cm]{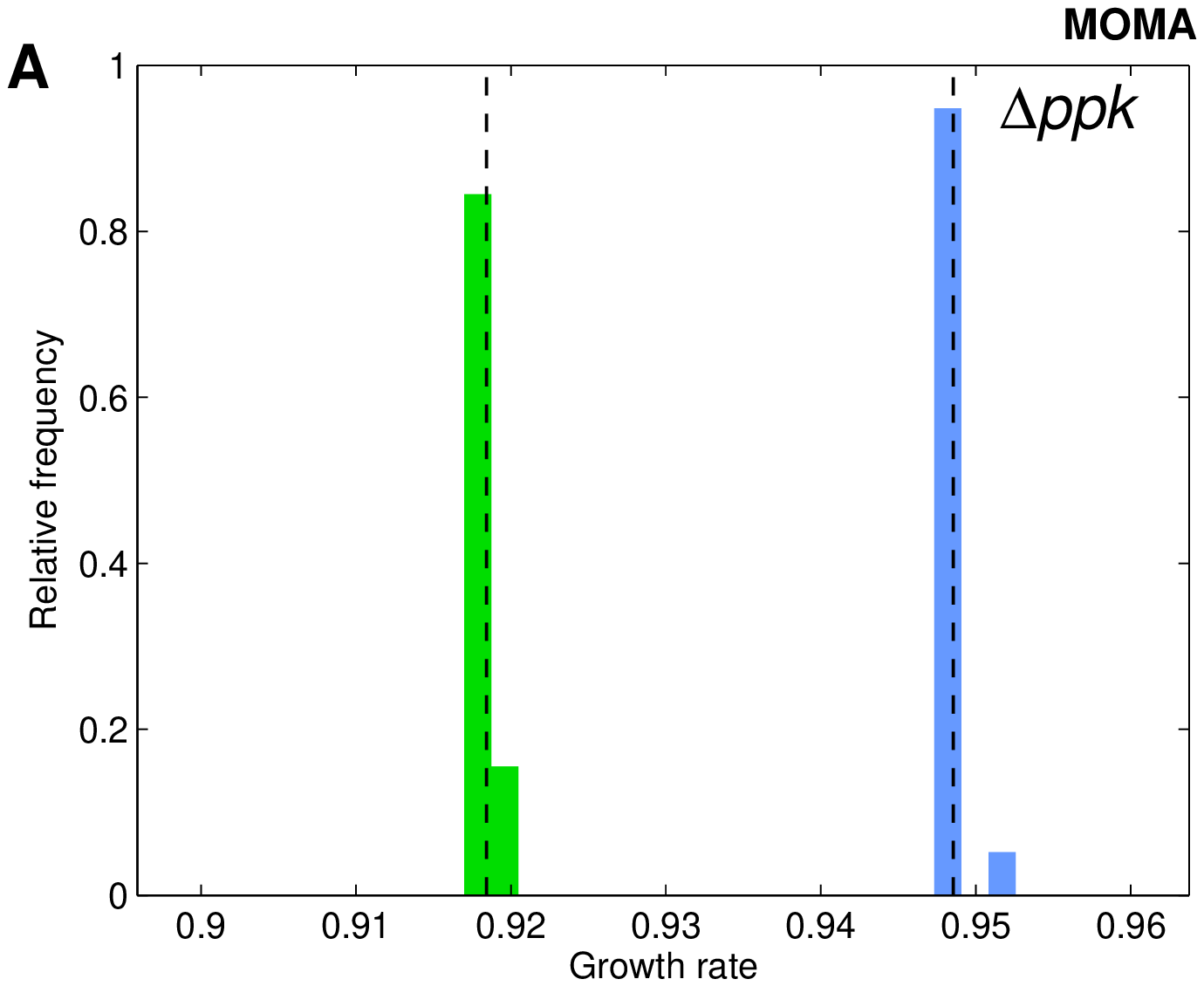} &
\includegraphics[angle=0,width=8.0cm]{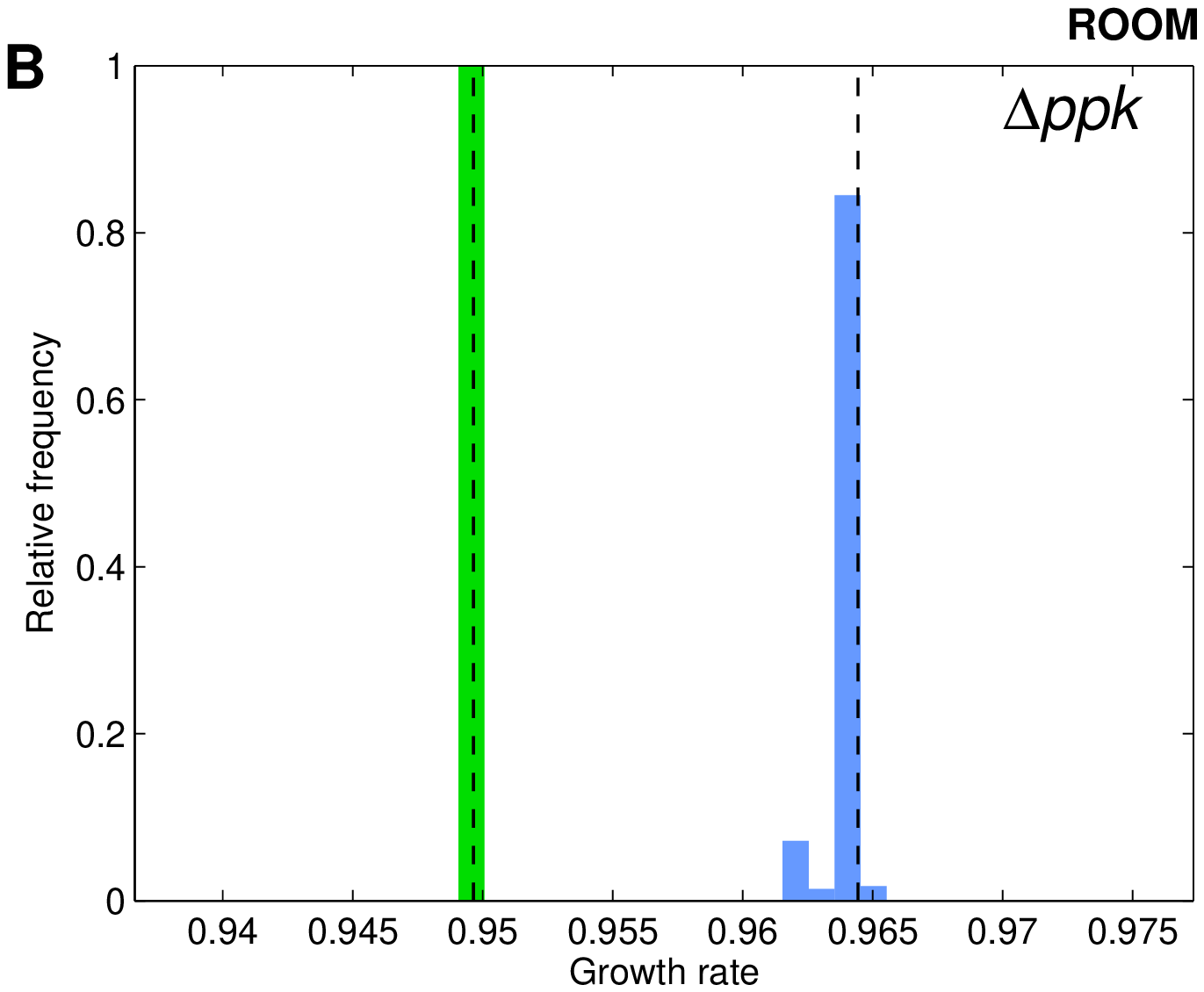} \\
\includegraphics[angle=0,width=8.0cm]{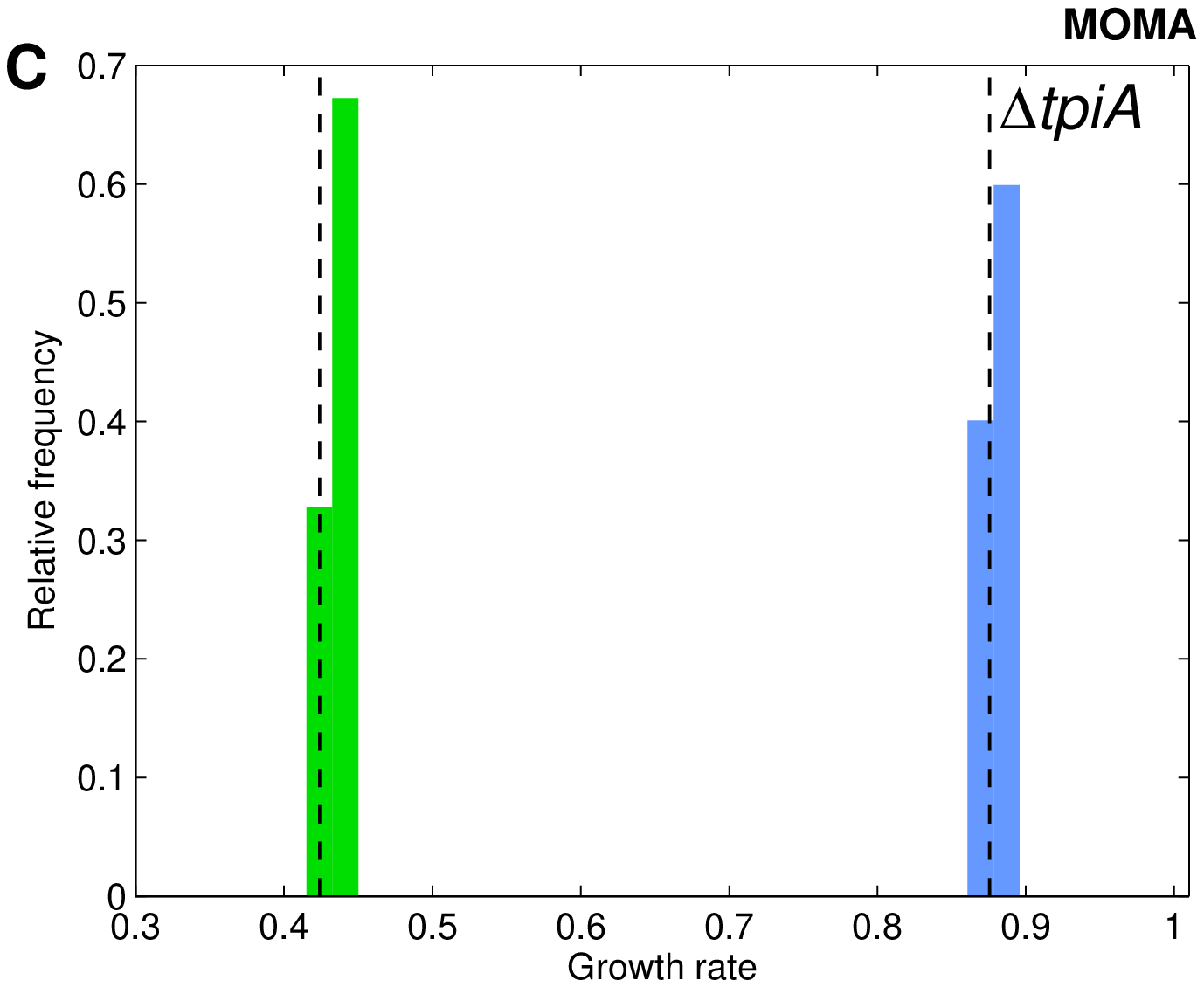} &
\includegraphics[angle=0,width=8.0cm]{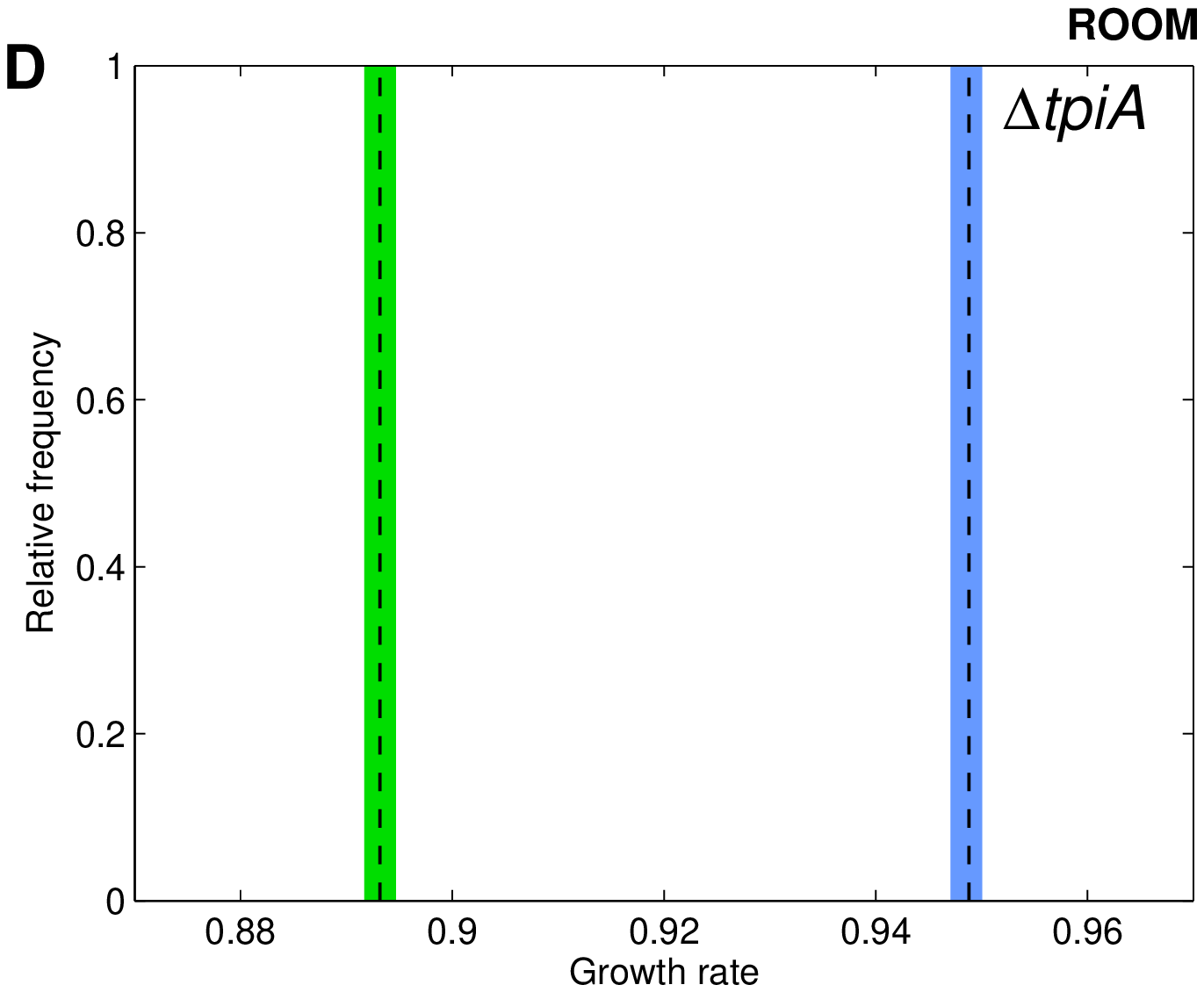}
\end{array}$
\end{centering}
\caption{Impact of alternate optima. ({\capitfont A} and {\capitfont B}) Sensitivity to alternate optima
for the {\capitfont ppk}-knockout perturbation according to MOMA ({\capitfont A}) and ROOM
({\capitfont B}) (based on 12,760 different pairs of optimal states). ({\capitfont C} and {\capitfont
D}) Sensitivity to alternate optima for the {\capitfont tpiA}-knockout perturbation according to
MOMA ({\capitfont C}) and ROOM ({\capitfont D}) (based on 20,880 different pairs of optimal
states).  Each panel shows the probability distributions of the predicted early
postperturbation growth rates for strains in  which the latent pathways are
available (green) and disabled (blue). All growth rates are normalized by the
optimal growth rate of the wild type. Dashed lines indicate the growth rate
predictions presented in the results of the main text. In all cases, the growth rates 
are tightly clustered around these values, and the two distributions do not overlap.}
\label{figs1} 
\end{figure*}

\pagebreak
\begin{figure*}[!ht]
\centering
\includegraphics[angle=0,width=8.0cm]{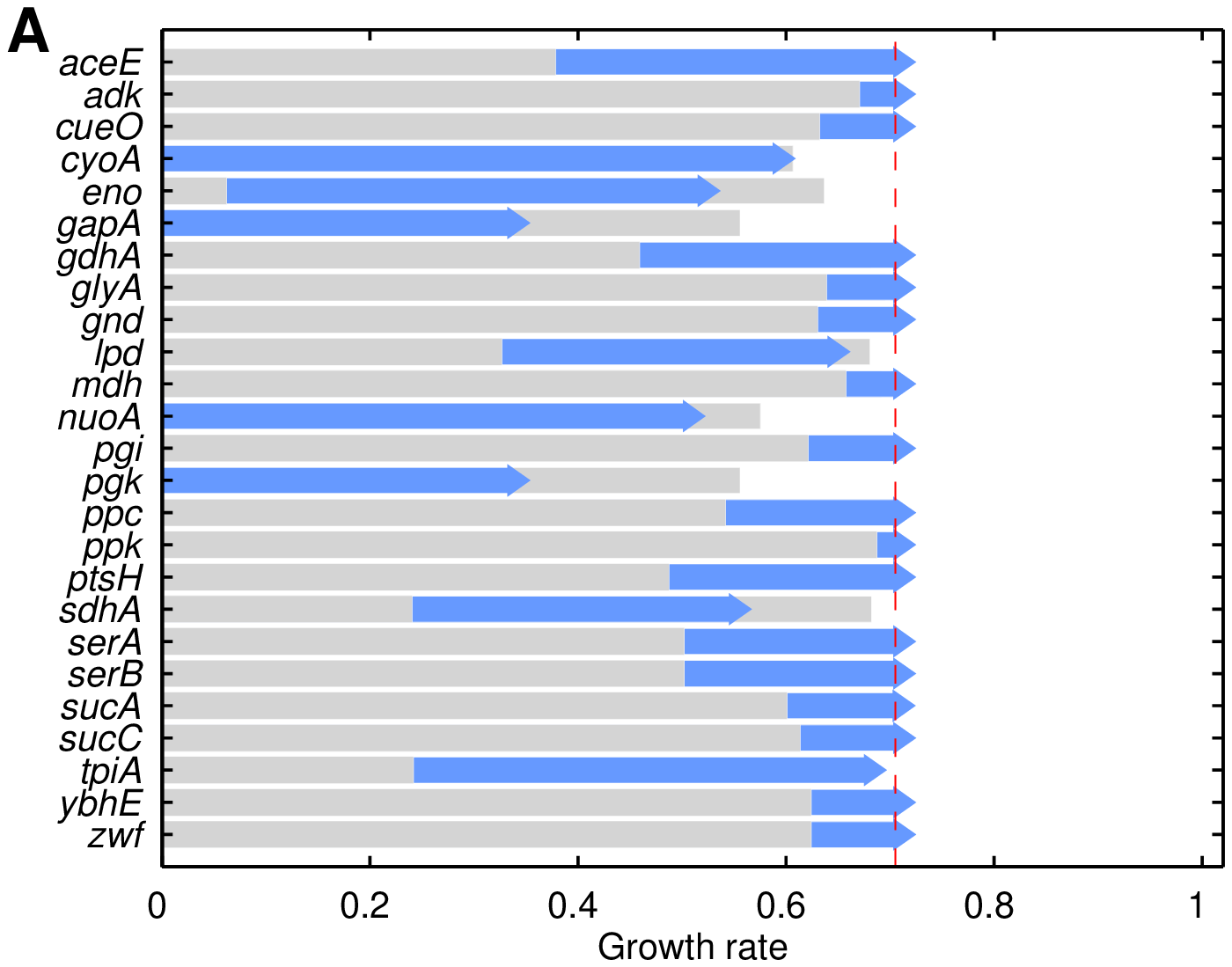} \\
\includegraphics[angle=0,width=8.0cm]{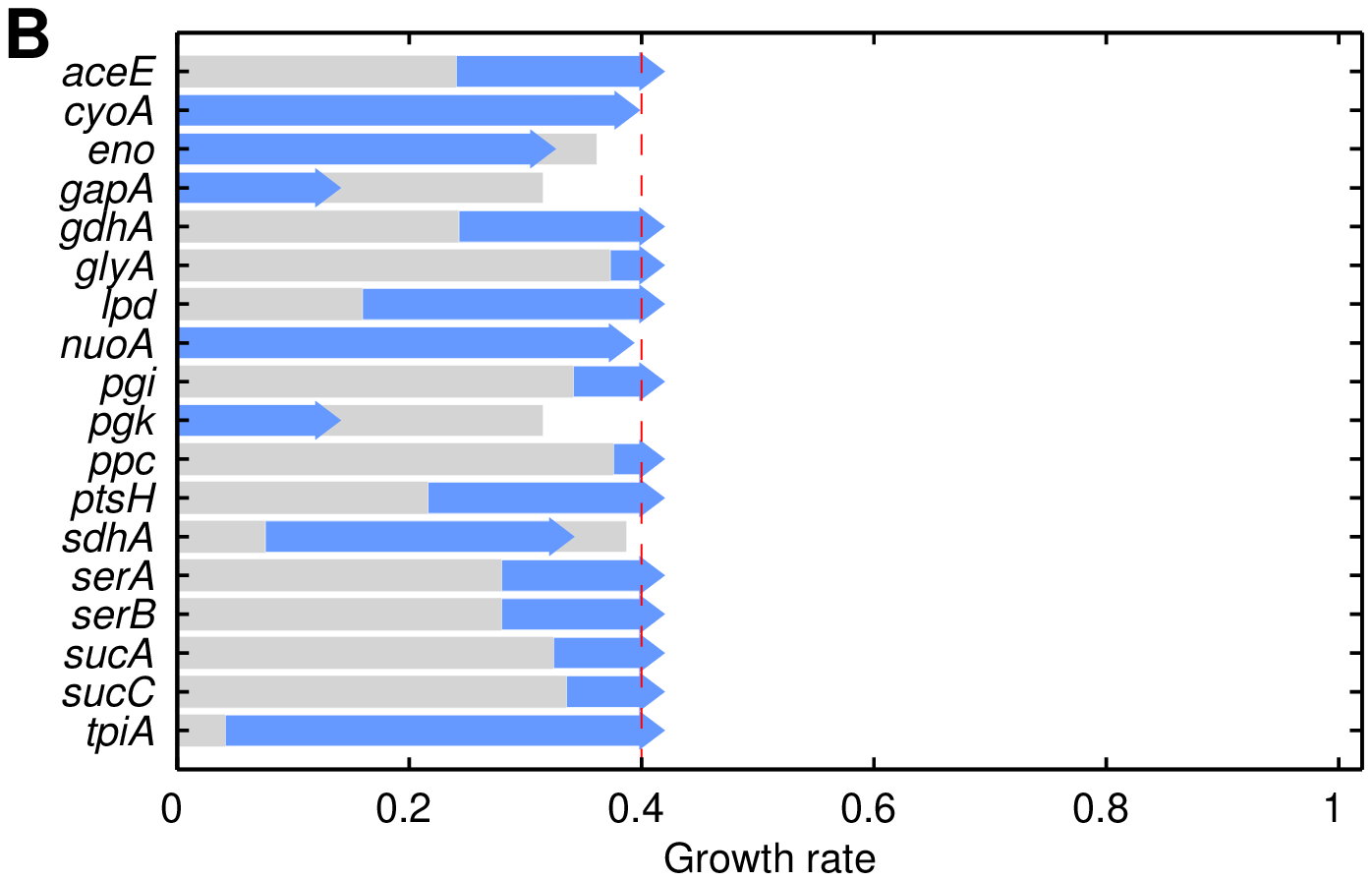}
\caption{Impact of nonoptimal reference states. ({\capitfont A} and {\capitfont B}) Effect of latent
pathway upregulation using reference metabolic states (wild type and ``adapted''
mutant) whose growth rates are limited to 70\% ({\capitfont A}) and 40\% ({\capitfont B}) of
the respective optima. For each knockout, the panels show the change in the
MOMA-predicted growth rate after the perturbation when we limit the magnitude
of the flux through each latent reaction to the corresponding highest value between
the two reference states. The dotted lines and gray bars indicate the maximum
allowed growth rates for the wild type and knockout mutants, respectively. 
All values are normalized as a fraction of the maximum wild-type
growth rate. We only display cases that show a significant difference in growth
rate ($>$ 1\% of the wild-type optimum) when the latent pathways are
downregulated. }
\label{figs2} 
\end{figure*}

\begin{figure*}[!ht]
\includegraphics[angle=0,width=8.0cm]{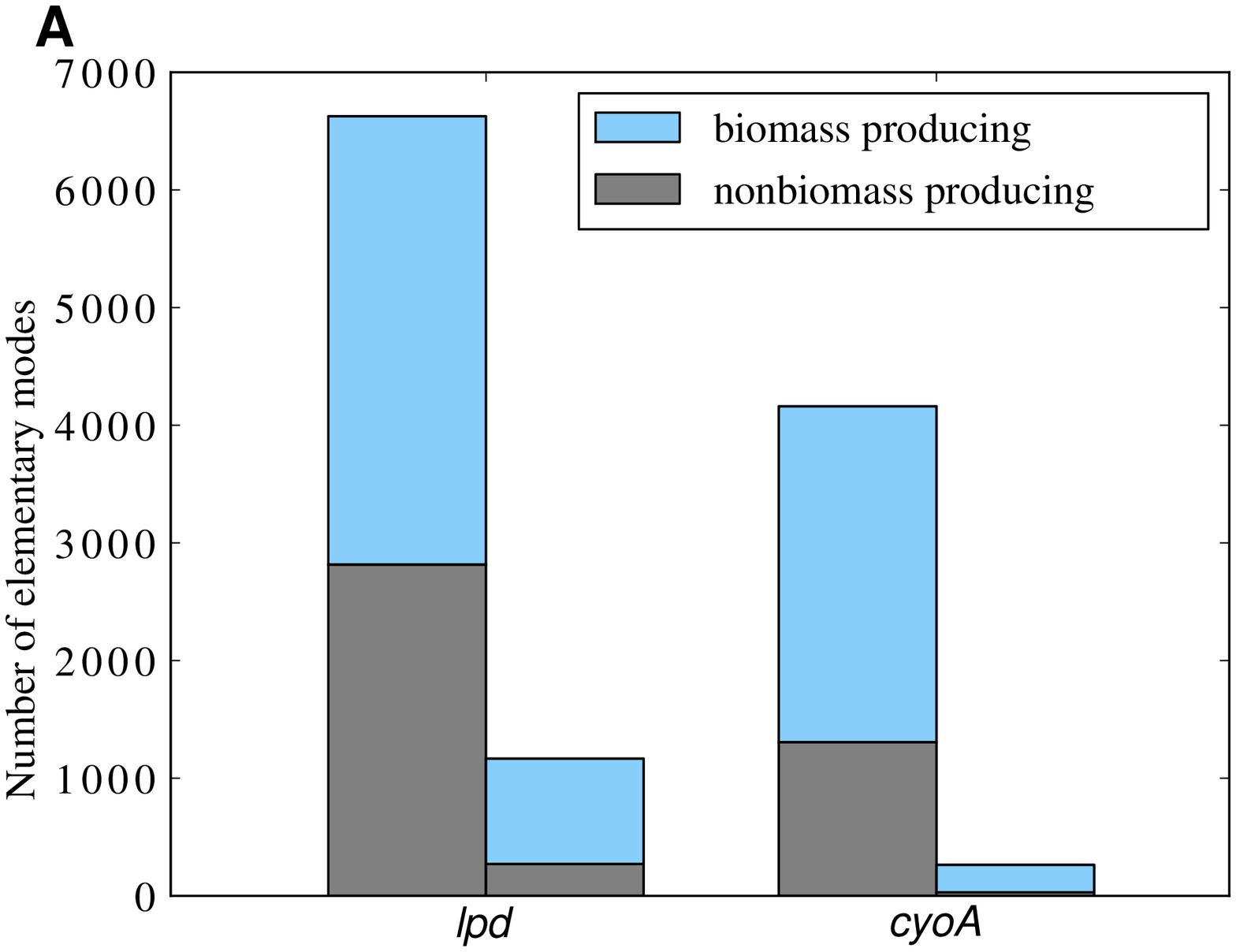}
\includegraphics[angle=0,width=8.0cm]{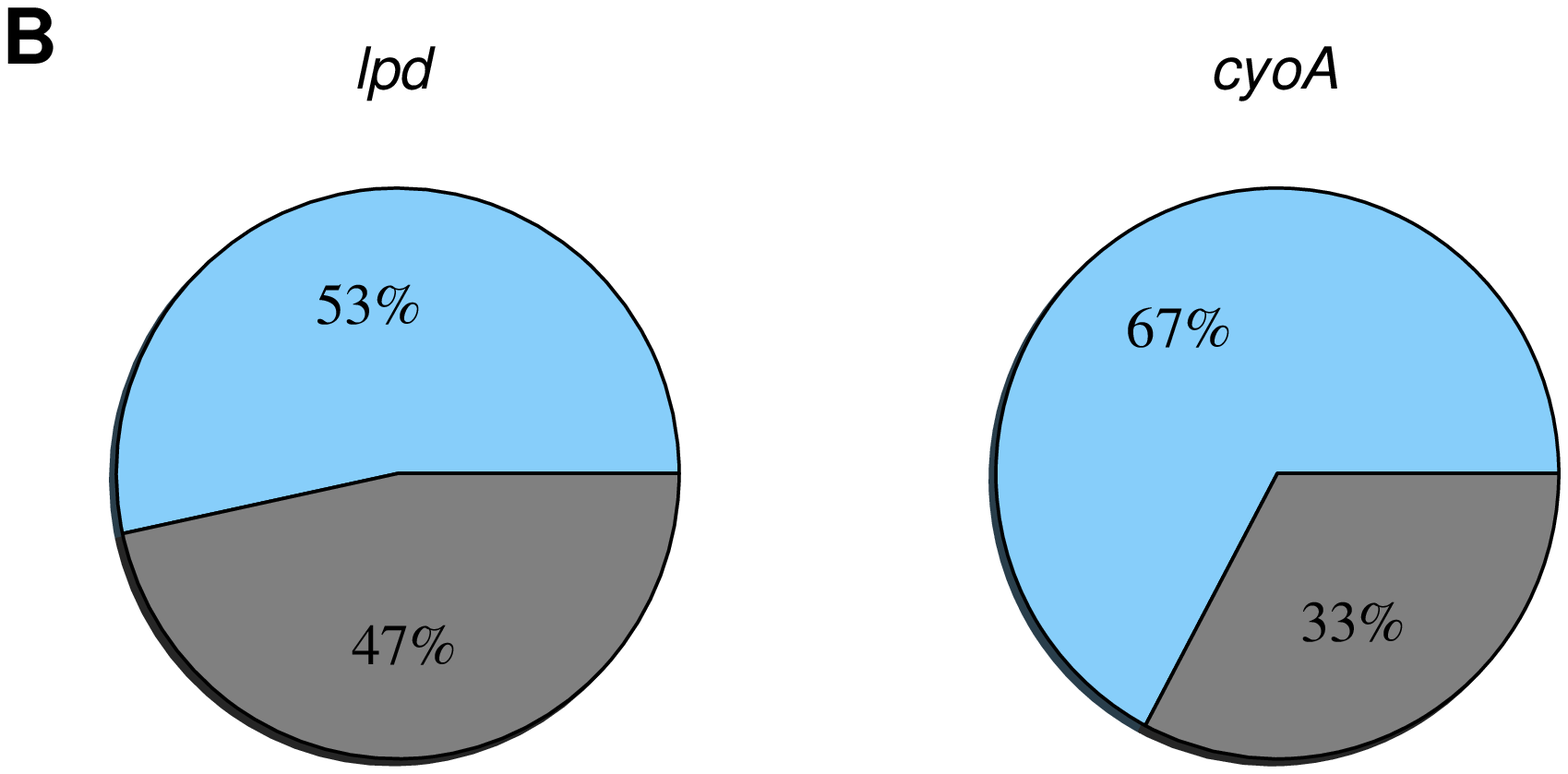}
\caption{Effect of latent pathway removal on elementary modes of {\capitfont E. coli}'s
central metabolism. These simulations are based on the {\capitfont lpd-} and {\capitfont
cyoA-}knockout mutants. ({\capitfont A}) Number of elementary modes having a zero
(gray) and positive (blue) biomass component, before (left bar, each mutant)
and after (right bar, each mutant) the removal of latent pathways predicted by
MOMA. ({\capitfont B}) Percentage of elementary modes of each type disabled by latent
pathway removal. Latent pathway removal renders a large number of elementary
modes unavailable for use. Strikingly, the majority of the eliminated modes involve
biomass production.}
\label{elementary-mode-analysis}
\end{figure*}

\pagebreak
\begin{figure*}[!ht]
\centering
\includegraphics[angle=0,width=16.0cm]{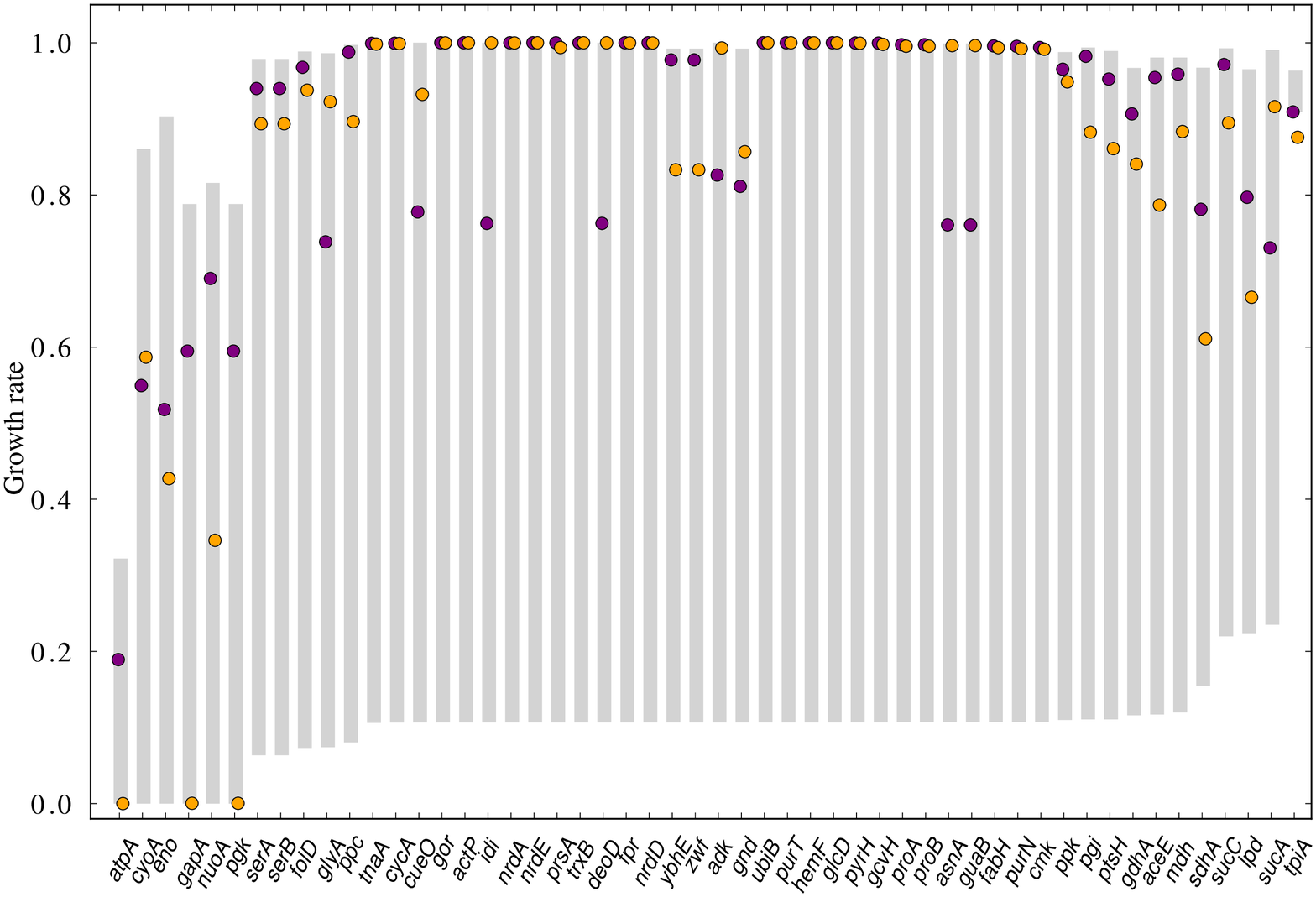}
\caption{Range of possible growth rates when all optimally-inactive reactions
are removed. The gray bars correspond to the range of realizable growth rates in
each of the 52 knockout mutants when all reactions not active in the
corresponding growth-maximizing state of the mutant are disabled. Purple circles
correspond to the MOMA-predicted growth rate in response to each knockout in
this case. Orange circles correspond to the MOMA-predicted growth when only the
latent pathways are removed, as presented in the main text. Although this
extreme pathway removal is also predicted to increase growth following a
knockout perturbation, low-growth states remain available for the large majority
of mutants. All growth rates are normalized by that of the optimal wild type.}
\label{extreme-removal}
\end{figure*}

\pagebreak
\setcounter{table}{0}
\begin{table*}[t] 
\caption{Comparison between linear ROOM and integer ROOM.}
\begin{center}
\begin{tabular*}{0.75\textwidth}{@{\extracolsep{\fill}} cccc} 
& & Integer ROOM & \\
\hline
 & + & = & $-$ \\
\hline
Linear ROOM & & & \\
+ & 12 & 0 & 0\\
=  & 12 & 27 & 1\\
$-$  & 0 & 0 & 0 \\
\hline
\end{tabular*}
\begin{tabular*}{0.75\textwidth}{@{\extracolsep{\fill}} l}
\begin{minipage}[t]{0.75\textwidth}%
\noindent The rows and columns indicate the number of single-gene knockout
perturbations for which the early post-perturbation growth-rate is larger ($+$),
smaller ($-$), or negligibly different ($=$) when the latent pathways are
disabled beforehand compared to the case in which they are available. As in the
main text, we define a change to be negligible if its magnitude is less than 1\%
of the wild-type growth rate. The majority of cases  are on the diagonal,
indicating that the two methods predict the same growth impact of latent pathway
removal. There are no mutants for which the two methods predict a significant
change in opposite directions.
\end{minipage}
\end{tabular*}
\end{center}
\label{tables1}
\end{table*}

\begin{table*}[h!]
\caption{Summary of the predicted impact of latent pathways under
nutrient-specific regulatory constraints.}
\begin{tabular*}{0.95\textwidth}{@{\extracolsep{\fill}} lccc }
& MOMA & ROOM & Random\\
\hline
Latent reactions for individual perturbations: & &\\
\;\;
All knockout perturbations & +10.5\, (13.2)\%~ & ~+1.5\, (3.8)\%~ & ~+61.0\,
(9.4)\% \\
\;\;
Significant differences \tablenote{By more than 1\% of the wild-type growth rate: 57\% (MOMA), 26\%
(ROOM), and 100\% (random) of the perturbations.} & +18.4\, (12.8)\%~ & ~+5.5\, (5.9)\%~ &
~+61.0\, (9.4)\% \\
\;\;
Number of reactions removed & 267\, (82) & 96\, (45) & 902\, (5) \\ 
\hline
Simultaneously nonessential latent reactions: & & \\
\;\;
All knockout perturbations  & +8.8\, (11.2)\%~ & ~+1.5\, (3.8)\%~ &  \\
\;\;
Significant differences \tablenote{By more than 1\% of the wild-type growth rate: 52\% (MOMA) and 24\%
(ROOM) of the perturbations.} & +16.7\, (10.4)\%~ & ~+5.8\, (5.9)\%~ &
\\
\;\;
Number of reactions removed & 235\, (76) & 82\, (38) & \\ 
\hline
\end{tabular*}
\begin{tabular*}{0.95\textwidth}{@{\extracolsep{\fill}} l}
\begin{minipage}[t]{0.95\textwidth}%
Corresponds to Table 1 of the main text for 46 knockout mutants in a modified
{\tabtextifont i}AF1260 {\tabtextifont E. coli} model in which 152 reactions have been disabled to
account for known regulatory constraints in aerobic glucose medium conditions.
\end{minipage}
\end{tabular*}
\label{tables2}
\end{table*}

\end{document}